\documentclass[letterpaper,twocolumn,10pt]{article}
\usepackage{usenix-2020-09}

\usepackage{tikz}
\usepackage{amsmath}

\usepackage{breakurl}
\hypersetup{breaklinks=true}

\usepackage{url}
\usepackage[english]{babel}
\usepackage{blindtext}
\usepackage{dirtytalk}
\usepackage{fancyvrb}
\usepackage{algorithm}% http://ctan.org/pkg/algorithms
\usepackage[noend]{algpseudocode}
\usepackage{varwidth}% http://ctan.org/pkg/varwidth
\usepackage{epigraph}
\usepackage{authblk}

\usepackage[capitalize, noabbrev]{cleveref}
\usepackage{pifont}% http://ctan.org/pkg/pifont
\usepackage{subcaption}

\usepackage{multirow}

\usepackage{pgfplots}
\usetikzlibrary{pgfplots.groupplots}
\usepackage{filecontents}
\pgfplotsset{compat=1.16}
\usepgfplotslibrary{groupplots}
\usepgfplotslibrary{fillbetween}
\usetikzlibrary{patterns}

\usepackage{enumitem}

\newcommand{\methodology}{\textbf{Methodology.}}

\newcommand*\samethanks[1][\value{footnote}]{\footnotemark[#1]}

\definecolor{cambridgeblue}{rgb}{0.64, 0.76, 0.68}
\definecolor{violet}{RGB}{96, 30, 169}
\long\def\com#1{}

\long\def\todo#1 {}
\long\def\cb#1{}
\long\def\baf#1{}
\long\def\gf#1{}
\long\def\xxx#1{}
\long\def\old#1{}
\long\def\veg#1{}
\long\def\cey#1{}

\def\name{Limix\xspace}
\def\Name{Limix\xspace}
\def\eg{e.g.,\xspace}
\def\ie{i.e.,\xspace}
\def\crdb{CockroachDB\xspace}

\algnewcommand\algorithmicinput{\textbf{I:}}

\algblockdefx[StructExt]{StructExt}{End}%
[2][]{\textbf{struct} \textsc{#1} \textbf{extends} \textsc{#2}}%
[0]{}

\algblockdefx[Struct]{Struct}{End}%
[1][]{\textbf{struct} \textsc{#1}}%
[0]{}

\algblockdefx[DoParallel]{DoParallel}{End}
[0]{\textbf{do in parallel}}
[0]{}

\algrenewcommand{\algorithmiccomment}[1]{\hskip3em$\rightarrow$ #1}
\algrenewcommand{\algorithmiccomment}[1]{\textcolor{gray}{\textbf{$//$ #1}}}
\algrenewcommand{\algorithmiccomment}[1]{\textcolor{gray}{ $//$ #1 }}

\usepackage[compact]{titlesec}
\titlespacing{\section}{0pt}{2ex}{1ex}
\titlespacing{\subsection}{0pt}{1ex}{0ex}
\titlespacing{\subsubsection}{0pt}{0.5ex}{0ex}

\begin{document}
%-------------------------------------------------------------------------------

%don't want date printed
\date{}

%\title{\Large \bf Coordinating Distributed Systems While Limiting Lamport
%              Exposure to Distant Failures}
\title{\Large \bf Limiting Lamport Exposure to Distant Failures \\
		in Globally-Managed Distributed Systems }

%for single author (just remove % characters)
\author[1]{
{\rm Cristina B\u{a}sescu}%
}
\author[1]{
{\rm Georgia Fragkouli}%
}
\author[1]{
{\rm Enis Ceyhun Alp}%
}
\author[2]{
{\rm Michael F. Nowlan\thanks{Nowlan's and Faleiro's work conducted while affiliated with Yale University, New Haven, CT, USA.} }
}
\author[3]{
{\rm Jose M. Faleiro\samethanks{} }
}
\author[4]{
{\rm Gaylor Bosson\thanks{Bosson's and Cong's work conducted while affiliated with Ecole Polytechnique F\'{e}d\'{e}rale de Lausanne (EPFL), Switzerland.} }
}
\author[5]{
{\rm Kelong Cong\samethanks{} }
}
\author[1]{
{\rm Pierluca Bors\`{o}-Tan}%
}
\author[1]{
{\rm Vero Estrada-Gali\~{n}anes}%
}
\author[1]{
{\rm Bryan Ford}%
} % end author

\affil[1]{Ecole Polytechnique F\'{e}d\'{e}rale de Lausanne (EPFL), Switzerland}
\affil[2]{Reflect, Philadelphia, PA, USA}
\affil[3]{Microsoft Research, USA}
\affil[4]{Taurus Group SA, Geneva, Switzerland}
\affil[5]{KU Leuven, Belgium}

\com{
\author{
{\rm Cristina B\u{a}sescu}\\
Swiss Federal Institute of Technology (EPFL)
\and
{\rm Georgia Fragkouli}\\
Swiss Federal Institute of Technology (EPFL)
% copy the following lines to add more authors
\and
{\rm Enis Ceyhun Alp}\\
Swiss Federal Institute of Technology (EPFL)
\and 
{\rm Vero Estrada-Gali\~{n}anes}\\
Swiss Federal Institute of Technology (EPFL)
\and 
{\rm Michael F. Nowlan}\\
Reflect
\and 
{\rm Jose M. Faleiro}\\
Microsoft Research
\and 
{\rm Gaylor Bosson}\\
Taurus Group SA
\and 
{\rm Kelong Cong}\\
KU Leuven
\and 
{\rm Pierluca Bors\`{o}-Tan}\\
Swiss Federal Institute of Technology (EPFL)
\and 
{\rm Bryan Ford}\\
Swiss Federal Institute of Technology (EPFL)
} % end author
}
\maketitle

%\epigraph{A distributed system is one in which the failure of a computer
%you didn't even know existed can render your own computer unusable.}%
%{Leslie Lamport}

%\epigraph{All problems in computer science can be solved by another level of indirection, except for the problem of too many layers of indirection.}%
%{David Wheeler}

\begin{epigraphs}

\qitem{A distributed system is one in which the failure of a computer
you didn't even know existed can render your own computer unusable.}%
{Leslie Lamport}

\qitem{All problems in computer science can be solved by another level of indirection, except for the problem of too many layers of indirection.}%
{David Wheeler}

\end{epigraphs}

\begin{abstract}

Globalized computing infrastructures offer the convenience and elasticity of
globally managed objects and services,
but lack the resilience to distant failures that localized
infrastructures such as private clouds provide.
Providing \emph{both} global management and resilience to
distant failures, however, poses a fundamental problem
for configuration services:
How to discover a possibly migratory, strongly-consistent service/object
in a globalized infrastructure without dependencies on globalized state? 
\com{
How can we discover a possibly migratory strongly-consistent object in a globalized infrastructure,
without d on global configuration?
The crux of providing \emph{both} global management and resilience to
distant failures entails a fundamental indirection conundrum. 
A possibly migratory object in a globalized infrastructure cannot
be addressed simply by a static name, but requires another ``level of indirection''.  
}%
\name is the first metadata configuration service that addresses this  
problem.
With \name, global strongly-consistent data-plane services and objects are
insulated from remote gray failures
\com{
failures, network partitions, and misconfiguration,}%
by ensuring that the definitive, strongly-consistent metadata for any object is
always confined to the same region as the object itself. 
\name guarantees availability bounds: any user can continue accessing any strongly consistent object that
matters to the user located at distance $\Delta$ away,
insulated from failures outside a small multiple of $\Delta$.
\com{  
\name guarantees availability with tight for any user accessing objects in any locality
by placing metadata nearby, with low overheads.
\cb{Find a better characterization of overheads.}
}%
We built a \name metadata service based on \crdb.
Our experiments on Internet-like networks and on AWS, using realistic trace-driven workloads,
show that \name enables global management and significantly improves availability over the state-of-the-art.

\end{abstract}

%We target a multi-DC deployment. 

\section{Introduction}

\com{
\cb{I think the new intro nicely frames the problem!
Somewhere in the introduction I suggest we make the point that many services already migrate
the data of interest close to users. \name ensures the users can keep accessing such data that interests
them by shielding the configuration service from remote failures.}
}

Organizations today face a choice between
\emph{localized} and \emph{globalized} computing infrastructure,
each alternative carrying important tradeoffs.
Localized infrastructure hosted at the organization's own site(s),
such as private clouds,
carry higher internal management burdens but offer greater local autonomy,
resilience to distant failures beyond the organization's control,
and can be necessary
to satisfy data privacy or digital sovereignty concerns.
Globalized infrastructure such as public clouds,
in contrast, offer many global management benefits:
\eg the convenience of instantiating objects or services on demand
without worrying about their location,
maximum elasticity in provisioning and adapting to changes in load,
and the ability to migrate existing data and services
without having to interrupt access or change their names.

\begin{figure}[!t]
\center
\includegraphics[width=0.95\linewidth]{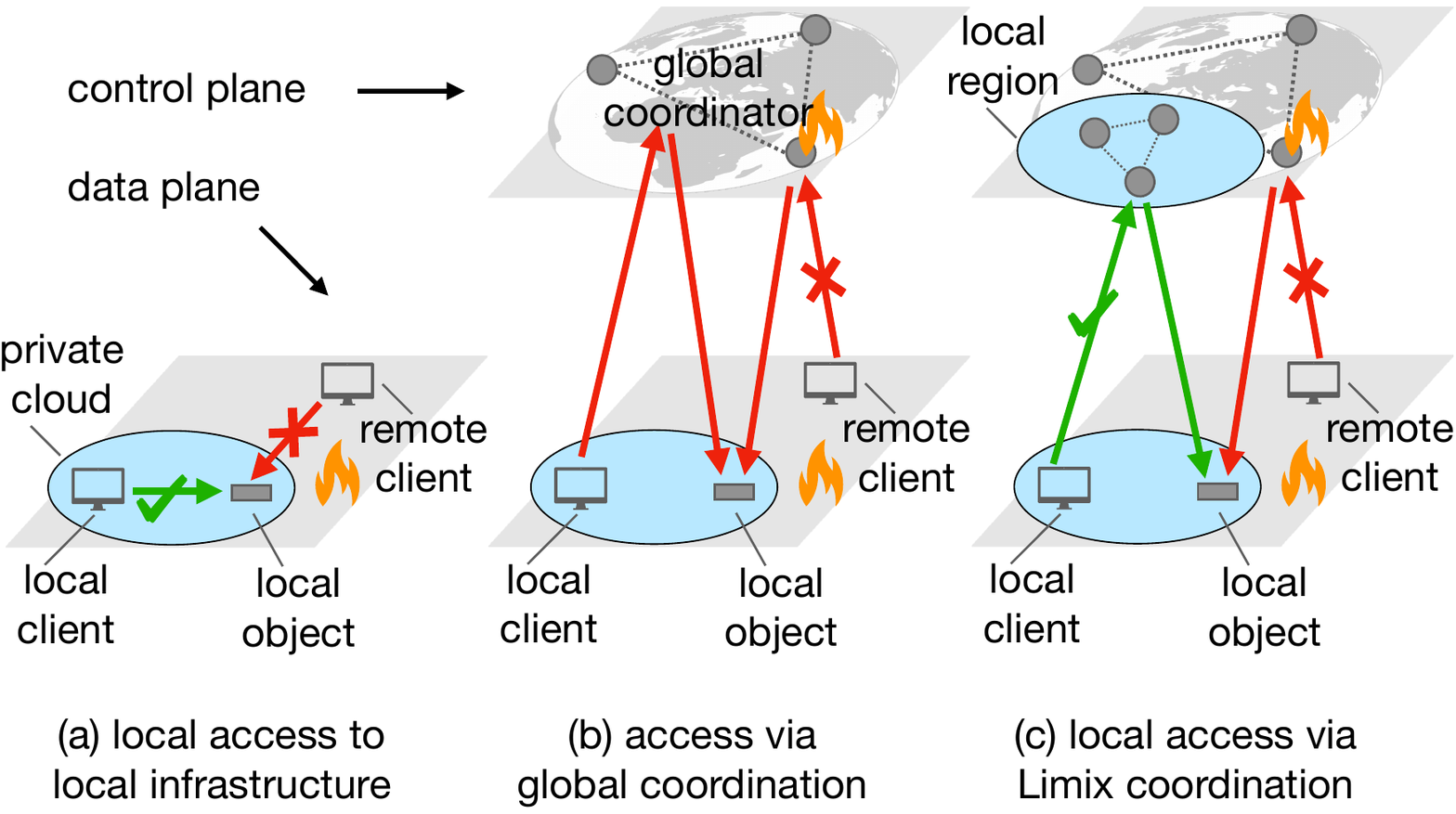}
\vspace{-0.6cm}
\caption{Lamport exposure in traditional
(a) local or (b) global infrastructure, and
(c) global infrastructure with Limix.
}
\label{fig:localization}
\vspace{-0.7cm}
\end{figure}

% \begin{figure}[!t]
% \center
% \includegraphics[width=0.95\linewidth]{figures/localization.eps}
% \caption{Lamport exposure in traditional
% (a) local or (b) global infrastructure, and
% (c) global infrastructure with Limix.
% \baf{It might be worth trying to replace the big "globe" circles
% in the figure with an \emph{extremely} light-colored (light-grey-on-white)
% "globe" clip-art to make it clearer what it's supposed to represent.
% But not sure because that risks making the figure too busy.}
% }
% \label{fig:localization}
% %\vspace{-0.6cm}
% \end{figure}

Is it possible to achieve the local autonomy, failure resilience,
and digital sovereignty benefits of localized infrastructure,
together with the global management benefits of today's public clouds?
%Achieving such a ``best of both worlds'' appears fundamentally hard,
Achieving the best of both worlds appears fundamentally hard,
in part because this choice boils down to a basic indirection conundrum.
For clients to find and access localized infrastructure
such as private cloud services,
we can simply embed a locally-scoped or already
%``resolved'' location
resolved location
directly into the identifiers
of objects and services being accessed. 
%\com{network location information}
\com{
\baf{FYI, never put \textbackslash com\{\} blocks by themselves on a line!
	Doing this screws up the spacing by producing two (real) spaces
	instead of one -- look closely at the spacing between the words
	"location" and "directly" (in the pre-conflict text).
	Always start an inline \textbackslash com\{\} block
	immediately after the previous non-commented text
	(e.g., "location\textbackslash com\{\}")
	or else add a comment character \% after the end of the block
	to consume the following newline
	(e.g., "\textbackslash com\{ ... \}\%"). }
\cb{I'm sorry, I didn't realize. I'll follow the practices you described from now on.}
}%
%\com{(\eg URLs)}
%\baf{ "identifiers" instead of "names"? }
%\cb{I initially thought of using URIs, URNs and URLs to make the point, but decided
%against it, because it seems too much to explain so early in the intro.
%However, "identifiers" would be just right in that sense, because they contain
%location + name. I'll make the change.} 
Accesses by local clients to local objects can be simple, robust,
and insulated from remote failures or network partitions
outside the relevant domain,
as illustrated in \cref{fig:localization}(a).
But these objects are then ``fixed''
and cannot be migrated or managed globally
without changing their names.
%\cb{The term Lamport exposure only appears in 
%\cref{fig:localization}
%in the entire intro. Do we need to define it in the text,
%even if it is one sentence?}

Global management, in contrast,
requires ``another level of indirection'':
typically a distributed coordination or metadata service,
allowing clients located anywhere to discover the location and status
of any object or service of interest.
\com{
This coordination service enables an object's location, access policies,
and other applicable metadata to change without destroying the object
or changing its name,
thereby facilitating the elasticity, migration,
and other benefits of global management.
}
These services, however,
expose clients to global \emph{gray failures}~\cite{huang17gray} such as
network partitions~\cite{alquraan18cloudfailures,haeberlen05glacier,fastly-may21},
misconfigurations~\cite{arbor,wsj-awsoutage},
or cascading failures~\cite{cloudflare-nov20,brooker2020millions,alquraan18cloudfailures,fastly-may21},
far beyond of the organization's geographic locality or domain of control --
\ie exposure to ``the failure of a computer you didn't even know existed.''
This global failure exposure usually applies
even when both the client
and the target data or service are localized
to the same network or region
and have connectivity in the
underlying network~\cite{brooker2020millions,alquraan18cloudfailures}.
\cb{Not sure about the next two sentences.}Even if the target data
might be weakly consistent~\cite{li12gemini,dynamo,seredinschi16icg},
metadata is usually strongly consistent
for many reasons~\cite{cockroachdb,annamalai18akkio}
such as correct liveness determination, access control, and accounting.
Dependence on globally geo-replicated metadata, however,
can prevent even local clients from accessing local data
if a majority of metadata replicas fail or become unreachable,
as illustrated in \cref{fig:localization}(b).

%\begin{figure}[!t]
%\center
%\includegraphics[width=0.8\linewidth]{figures/limix_intro.pdf}
%\caption{Global management and resilience to remote partitions
%in different architectures. In contrast to existing cell-based architectures, \name
%provides resilience guarantees and better global manageability.}
%\com{
%Geo-replication and private clouds each fulfill only one
%of globalized management and resilience to remote partitions. \name fulfills both.
%\cb{Added this figure, we might want it or not.}
%}
%\label{fig:limix_intro}
%\vspace{-0.6cm}
%\end{figure}

To address this conundrum we introduce \name,
the first distributed coordination architecture
that enables global management while also guaranteeing
that localized accesses within a region of interest
are insulated from global failures beyond that region,
as illustrated in \cref{fig:localization}(c).
\com{
\cb{One could argue that Physalia provides similar properties. Physalia has other
points where it falls short, but I'm wondering if it's safe to claim we are the first.}
}
\Name ensures that the definitive, strongly-consistent metadata
for any data-plane object or service is always collocated in the same region
as the object itself.
Metadata in \name thus enjoys
a \emph{fate-sharing} relationship~\cite{clark88design}
with both the target object
and
with any local clients accessing the object from within the same region --
such as within an organization's own internal network,
or within a relevant geopolitical domain such as a country.
Metadata remains
\com{may still be}
strongly-consistent and geo-replicated,
but \name confines the definitive replicas
of an object's strongly-consistent metadata
to the same region as the object itself.

\com{
Additionally, 
if and when the data plane is affected
by remote correlated failures,
it is crucial that the configuration plane does not
suffer from the same correlated failures, otherwise it cannot keep up with the reconfiguration demands
from the control plane

\cb{I think fate-sharing could be misunderstood. I agree that metadata should be
in the same locality as the target objects. But, if and when the data plane is affected
by remote correlated failures,
it is crucial that the configuration plane does not
suffer from the same correlated failures, otherwise it cannot keep up with the reconfiguration demands
from the control plane, and everything crumbles.
Perhaps we could say that metadata should be in the same locality, but in an independent
deployment insulated from remote failures?}

Related work does not achieve these goals albeit under 
assumptions. Physalia~\cite{brooker2020millions}'s aims to limit the blast
radius of a failure, but does not provide strong availability guarantees
for localized user activity across zone boundaries, and its proprietary
zone definition obstructs insights into availability guarantees.
Also, reconfiguration in Physalia takes minutes, which might be suitable for object that migrate slowly, such as
volumes, but unsuitable for services that migrate small objects very
frequently.
}

Cell-based configuration services like Physalia~\cite{brooker2020millions}
improve failure resilience for users within the same cell.
These provider-managed cells, however, do not offer users
direct transparency into or control over each user's {\em Lamport exposure} --
the set of infrastructure components whose failure could affect the user --
as discussed later in \cref{sec:motivation:lamport}.
Further, cell-based configuration services offer no guarantees
for user activity that crosses cell boundaries.
%Finally, Physalia does not support global manageability for small objects that
%migrate frequently, because of slow reconfiguration.
%\com{
%\baf{ The further niggly details in the rest of this paragraph
%	definitely aren't fundamental enough to include in the intro:
%	they should be cut or moved somewhere else appropriate:
%Unfortunately, some of its design choices that are suitable for
%a block storage data plane cannot accommodate other applications.
%Applications with dynamic user
%interactions necessarily have many
%users interacting across cells, but would receive no formal guarantees. \cb{Maybe cite pando, spanstore?}
%Also, Physalia's reconfiguration takes on the
%order of minutes, which is suitable only for global manageability of object
%that migrate slowly, such as volumes, but unsuitable for services that migrate
%small objects frequently~\cite{annamalai18akkio}.  
%}%baf
%\cb{Addressed this comment. I juat added the point on global manageability, in a more concise form. It helps
%for the figure; Is that still too much?}
%}%
%\cref{fig:limix_intro} illustrates these points, comparing different architectures with respect to
%global manageability and resilience to remote failures.
%\com{
%\baf{ If you're going to include this figure,
%you need to explain it at least a bit better,
%either here or perhaps better, directly in the caption. }
%\cb{Addressed. Is the above better?}
%}%

\textbf{Challenges and Contributions.}
\name's design decisions for
practical global manageability and resilience become apparent when
we target applications where the locations from which data-plane items are accessed
change dynamically.
Data stores and locality management services with dynamic data access
locality already migrate strongly-consistent data close to users~\cite{cockroachdb,annamalai18akkio}.
\name ensures that users can continue
accessing such nearby items under remote failures even during migration, 
while preserving strongly-consistent access.

Constraining the placement of strongly-consistent metadata in localities
\com{this way, however,}
creates the further efficiency and scalability challenge
of enabling any clients outside an object's current region
to find the object without incurring the costs of either
replicating all location information proactively across all regions,
or potentially having to search all regions during any metadata query.
\Name builds on techniques from
compact graph summarization theory~\cite{thorup01approximate,thorup01compact}
to limit the bandwidth and processing costs of these global searches
to a small multiple of the baseline cost
of querying a single global metadata service.
The metadata-access costs of \name's failure insulation
is thus only about $2\times$
in the common case of an object administratively localized to a single region.
Since metadata query costs usually represent only a small fraction
of the total ``end-to-end'' costs of accessing most data-plane services,
a $2\times$ metadata query cost increase
is insignificant overall to most applications,
and is much lower than the $N\times$ metadata cost increase
that cell-based architectures with $N$ distributed discovery services
(or the proactive replication of location hints across all $N$ regions)
would otherwise incur.
\com{
that a na\"ive search through all $N$ regions 
(or the proactive replication of location hints across all $N$ regions)
would otherwise incur.
}
Further, \name ensures by design that not only availability
but also metadata access latencies observed by local clients
are insulated from global outages or slowdowns,
and that they closely reflect the best local communication latencies available
on the underlying network.

To evaluate \name's applicability and performance, we prototyped a \name
% and a Physalia-inspired	\baf{Limix is this paper's focus, not Physalia.}
configuration service for an exposure-limiting key/value store. For
metadata/configuration storage, our prototypes use
CockroachDB~\cite{cockroachdb}, a widely-used, strongly-consistent distributed
data store.  Our experiments running realistic workloads based on metropolitan
traffic traces on AWS,
and on a testbed simulating realistic scenarios,
show that \name outperforms
Physalia's availability during reconfigurations \cb{add by how much}
while providing strong availability guarantees. 
Our experiments further explore the
tradeoffs between \name's overheads and availability guarantees: at scale, the
dynamic load overhead is logarithmic in the number of nodes and network width.
% Summary of contributions                                                      

In summary, the contributions of this paper are:
(1)
%\begin{itemize}[noitemsep]
%\item 
The design of \name, the first distributed metadata coordination service
that enables global management
while protecting local accesses from distant failures.
%\item
(2) An autozoning scheme ensuring by design that
a user accessing any data at a distance $\Delta$ away
is protected from all failures
occurring beyond a small multiple of $\Delta$.
(3)
%\item 
A prototype implementation of \name and Physalia on top of \crdb
with a comparative evaluation.
%\item An analysis confirming \name's availability guarantees.
%\end{itemize}

\section{Background and Motivation}
\label{sec:motivation}

This section gives the necessary background for a strongly-consistent
configuration service like \name. We first explain that the CAP theorem imposes
restrictions on the available \name can achieve, and that \name does not
conflict with the CAP theorem when prioritizing availability of local accesses.
Second, we focus on \name choice to prioritize user-perceived availability, for
the data-plane objects that matter to users. Reducing the Lamport exposure,
which is the user-centric viewpoint of \name, contrasts to Physalia's
provider-centric viewpoint, \ie blast radius. \gf{Swap the previous "first" and "second" points to match cirrent structure of section (lamport-vs-blast before CAP discussion)} Our discussions with
risk-sensitive customers share \name's viewpoint. Finally, we argue that
access locality is prevalent in globalized applications, and thus, \name's
focus on shielding local user activity leads to a sizeable increase in
user-perceived availability.

\com{
Finally, we estimate the prevalence of access locality.
% and the perceptions of risk-sensitive customers.
}

\subsection{Lamport exposure and blast radius}
\label{sec:motivation:lamport}

\com{
\baf{ I feel like this subsection should be moved up,
    before the last one: it's much more fundamental to this paper
    and gets to the main point much more quickly;
    the CAP theorem stuff would then add depth to that.}
\baf{ Also, the workshop paper's intro and abstract had some background text
    that I thought was pretty clear,
    including a fairly clear setup and definition of Lamport exposure --
    maybe some of that should be resurrected towards the beginning
    of the Background section? }
}

\begin{figure}[!t]
\center
\includegraphics[trim=0 116 0 145,clip,width=0.95\linewidth]{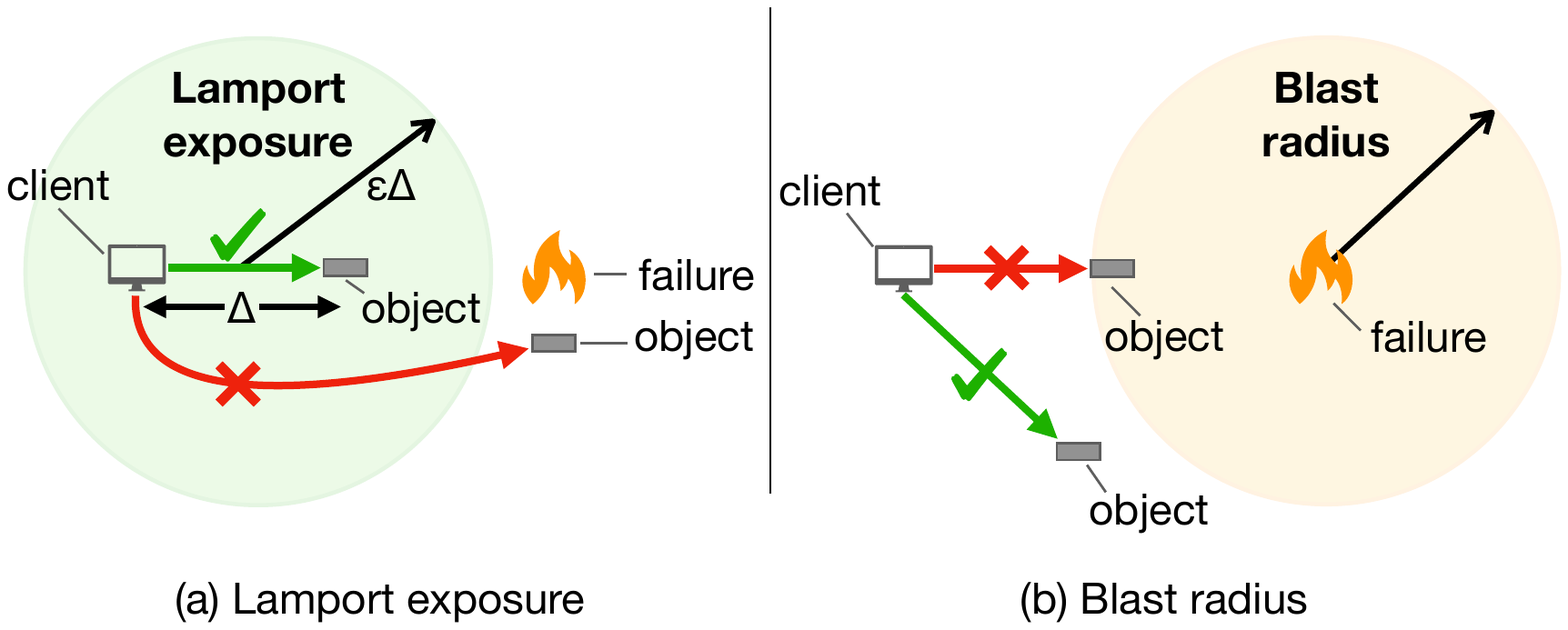}
\vspace{-0.4cm}
\caption{Illustration of Lamport exposure and blast radius.}
\label{fig:lamport-vs-blast}
\vspace{-0.4cm}
\end{figure}

\com{
\baf{ I think it would be better to present and explain Lamport exposure first,
then summarize and compare with Physalia and blast radius.
Keep this paper's contributions front and center. }
}

We informally define the \emph{Lamport exposure} of any given user $U$ as
the set of computing infrastructure -- 
\ie every ``computer $U$ didn't even know existed'' --
that ``can render $U$'s own computer unusable''~\cite{basescu21limix}.
Lamport exposure is thus meaningful only with respect to the activities
of some user $U$.

\Name seeks to place a strong bound or ``shield''
on the Lamport exposure of any user $U$ whenever $U$ accesses data or services
that are \emph{local}, \ie close to $U$ by any suitable distance metric.
We may define locality based on administrative boundaries
such as those of an organization or a country,
or via metrics such as round-trip time (RTT).
The availability of $U$'s local accesses should be unaffected
by remote failures and partitions
beyond \name's Lamport exposure shield:
not just by individual server failures but also by network partitions
and \emph{gray failures}, or partial failures
that can cascade into correlated failures and partial partitions.
Gray failures have a multitude of causes including
malfunctioning switches and software bugs that partially drop traffic
or prevent simplex
communication~\cite{alquraan18cloudfailures,cloudflare-nov20,haeberlen05glacier,fastly-may21}.
\Cref{fig:lamport-vs-blast}(a) illustrates a Lamport exposure bound
shielding a user's accesses to an object at distance $\Delta$
from failures beyond a distance $\epsilon \Delta$,
for some small factor $\epsilon \ge 1$
(ideally 1, but this may be unachievable).

\textbf{Locality matters.} By limiting Lamport exposure,
\name seeks to offer availability guarantees to users
for the data and services that matter most to them,
which are often local.
It is already common for data stores to migrate data close to
users to improve performance~\cite{cockroachdb,annamalai18akkio},
even without guaranteeing availability.
Privacy and sovereignty considerations often motivate
organizations or governments to require a user's ``data at rest''
to remain within the user's country, or a region such as the EU --
even if these policies cannot currently ensure that this data
will \emph{remain accessible} despite outages beyond the relevant borders.
Similar considerations motivate
many organizations to confine their most-critical data and applications
to local infrastructure such as private clouds,
giving up the benefits of global management.
In general,
people more willingly trust organizations and
services perceived to be more local~\cite{buss-trust,govt-trust}.
We hope that \name might enable providers to offer services to
more locality- and sovereignty-conscious users
who might currently avoid globally-managed infrastructure entirely.

In contrast to \name, Physalia~\cite{brooker2020millions} is a cell-based
architecture that aims for higher availability
by limiting the \emph{blast radius} of a failure. 
Blast radius represents the system components and objects
that could be affected by the propagation of a failure or partition
(\cref{fig:lamport-vs-blast}(b)). 
Blast radius is thus a complementary concept focused (or ``centered'')
on the location of a \emph{failure},
rather than on the location of a \emph{user} potentially affected by it.
From an infrastructure provider's perspective,
reducing blast radius can reduce the number of users affected by any
single failure. 
Being focused on the locations of failures
and heavily dependent on internal details of the provider's infrastructure,
however,
blast radius does not offer obvious guarantees
that appear directly meaningful to users.

\subsection{Coordination and the CAP theorem}

Strongly-consistent coordination systems (Zookeeper~\cite{hunt10zookeeper},
etcd~\cite{etcd}) are an essential building block for large scale distributed
applications. Distributed applications replicate their state in order to
enhance resiliency to failures, and to decrease latency through proximity to
clients. But an unsought side-effect is the need to coordinate these replicas.
Hence the need for coordination services: These provide a basic set of
operations -- such as liveness determination, correct identification of
the replicas storing a data item, lease holders, access control, accounting,
etc.  Because these functions need to be \emph{correct}, coordination services
implement consensus among the configuration replicas, ensuring
strong consistency of the configuration.

% CAP theorem
Coordination systems often are not critical to application availability until
failures occur.  Applications routinely bypass the coordination system for
configuration reads using leases~\cite{chubby, cockroachdb}.  \com{ Sometimes
reading merely eventually-consistent configuration stored in caches
suffices~\cite{hunt10zookeeper}.  }%
During partitions or failures, however,
the data plane cannot bypass the configuration system, because it needs to
reconfigure its data replicas.  This is when the configuration system becomes
critical to application availability and performance.  Reconfiguration requires
strongly-consistent configuration writes to agree on the new configuration.
Until reconfiguration completes, the data-plane may be partially (\eg operate
in read-only mode only) or fully unavailable. \cb{add apache downtime citation}

The requirements of configuration systems to be strongly consistent and available
seem to conflict with the CAP theorem~\cite{gilbert02cap}: under partitions, a
strongly-consistent (configuration) system cannot remain available (on both
sides of a partition).  However, \name does not conflict with the CAP theorem
when prioritizing availability of local accesses.  Remote users may not be able
to access remote data during partitions, but local users can, without breaking
strong consistency, and in many cases as described above, local data is what
interests users.  For this reason, \name collocates metadata with its data. 

To decrease the risk of being affected by remote gray failures, Brooker et
al.~\cite{brooker2020millions} suggest many smaller configuration service
deployments instead of a network-wide ``monolith''.  Deploying several
configuration services instead of a single globalized deployment is one of the
principles that \name also applies.  However, as opposed to deploying disjoint
cells, \name creates overlapping, redundant configuration service deployments,
organized to provide availability guarantees
for any user accessing any object or service.

\begin{figure}[!t]                                                              
\center                                                                         
%\vspace{-1.0cm}                                                                
\includegraphics[width=0.8\linewidth]{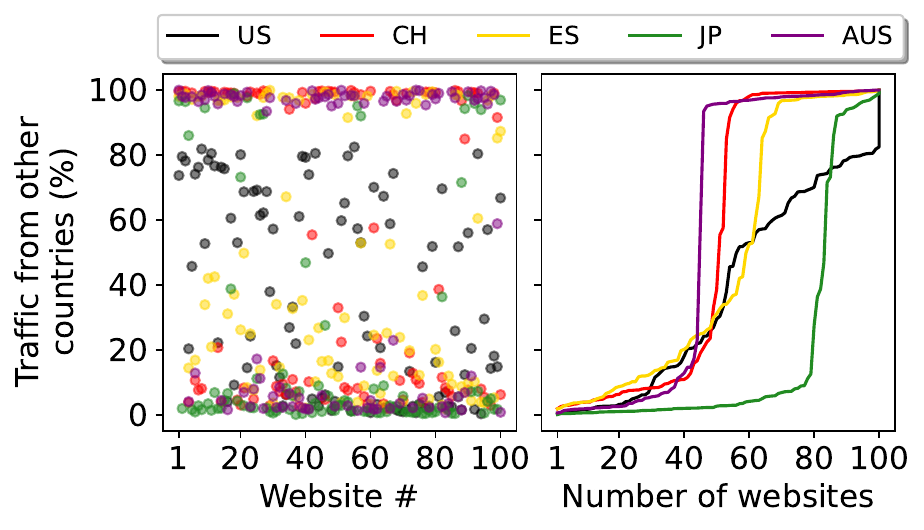}           
%\vspace{-1.5cm}                                                                
\caption{Prevalence of access locality. Top 100 websites in 5 countries and their traffic percentage
from other countries (individual websites and CDF).}
\label{fig:opportunity_top100}                                                  
\vspace{-0.4cm}                                                                
\end{figure}

\com{
Our concerns are different. Informed by our experience below, we take a         
\emph{user-centric} perspective:                                                
and we wish to limit any user's \emph{Lamport exposure}:                        
a guarantee that locally-provided services depend                               
only on local computing infrastructure,                                         
and that no failure ``of a computer you didn't even know existed'' --           
on the other side of the world,                                                 
or otherwise distant by some suitable metric --                                 
can ``render your own computer unusable.''~\cite{limix}. 
}

% interviews

% Notes from the "Jerry Gupta (Swiss Re)" interview, after listening to the recording
% - Almost everything is run on the cloud; more and more applications are moving from private to public clouds.
% - Outages are not going down; there are still lot of outages
% - Service sovereignty > blast radius
% - Companies want portability across clouds. Service sovereignty might enable portability if you replicate data, service, and applications. Should be treating clouds as bare-metal infrastructure. If service goes down on one cloud, should be able to move it to another cloud. Real-time portability and availability is the holy grail.
% - How do you minimize the probability of SLAs on availability not being met? Are they met? Probably not. Focus on technology evolution to meet SLAs.

%\subsection{Meaningful and enforceable availability}

\subsection{Perceptions of risk-sensitive customers}

Recent discussions we had with globalized infrastructure customers---a large
company and a non-profit organization, both international---informally
confirmed to us their need
for services that are globally-managed while providing availability guarantees
in the face of distant failures.  Applications increasingly move to globalized
infrastructures such as public clouds to benefit from the elasticity properties
and lower management effort. Despite the flexibility of including custom
clauses in the SLA, some customers are reluctant to trust these
infrastructures: Because the number of reported outages is still uncomfortably
high, they fear the reputation risk.  One customer suggested that a
good design that guarantees availability should recommend itself before getting
to the SLAs. Finally, customers would consider leveraging a service that limits
Lamport exposure -- one referring to this as ``the holy grail'' -- because its
guarantees are well-understood and more meaningful compared to simply limiting
the ``blast radius'' of a failure.

\subsection{Estimating prevalence of access locality}

\com{
Our hypothesis for \name's design is that many applications that have a strong local presence need
global manageability, as they also have a global presence.
\baf{ It's not clear how or if this particular hypothesis, as described,
	is directly supported by the data in the figures.
	Maybe the hypothesis should be something more like:
	"We observe that many applications and services
	exhibit bimodal access locality.
	A high proportion of accesses are by users
	mostly in a given country or other region
	representing the application's primary customer base or target audience.
	Other users of these same applications and services, however,
	tend to be globally distributed,
	accessing the service from anywhere (\eg roaming employees or expats).
	Thus, applications must efficiently support global accessibility
	for global users,
	while prioritizing maximum availability and performance
	for local users representing the most critical target customers.
	(paragraph)
	To test this bimodal-access hypothesis,
	we use the top 100 websites of several countries
	as rough proxies for applications,
	and examine the access distributions they exhibit.
	... }
}

We observe that many applications and services exhibit bimodal access locality.
A high proportion of accesses are by users mostly in a given country or other
region representing the application's primary customer base or target audience.
Other users of these same applications and services, however, tend to be
globally distributed, accessing the service from anywhere (\eg roaming
employees or expats).  Thus, applications must efficiently support global
accessibility for global users, while prioritizing maximum availability and
performance for local users representing the most critical target customers.

To test this bimodal-access hypothesis, we use the top 100 websites of five
OECD countries (namely United States (US), Switzerland (CH), Spain (ES), Japan
(JP) and Australia (AU)) as rough proxies for applications, and examine the
access distributions they exhibit.  Since these are the most visited websites
in their respective countries, we conjecture that they have a strong local
presence. In Figure~\ref{fig:opportunity_top100}, we show that these websites
also have a global presence, as 17--56\% (JP--AU) of the websites receive at
least $50\%$ of their traffic from outside of that country. The results of our
simple study does not make any assumptions about the consistency models used by
the websites, but it suggests that locally-prevalent applications are
indeed globally relevant as well.

\section{Setting and Goals}
\label{sec:setting}

In this section we discuss the main system components,
our assumptions and our goals.

\com{
\gf{The following paragraph can be omitted/moved to next section}
\name is the first distributed coordination architecture
that enables global management while also guaranteeing
that localized accesses within a region of interest
are insulated from global failures beyond that region.
We first define basic concepts, such as data plane items, sites and zones.
Then we introduce the challenges that \name needs to tackle: fulfilling data
consistency requirements, simultaneous item constraints\gf{not clear what are these constraints or why meeting them is challenging}, load balancing, item
migration and meaningful availability guarantees\gf{"meaningful" too abstract/unclear; replace with "user-centric, mathematical"? also, unclear why this is challenging (e.g., add something like "despite failures")}. 
}

\name involves the following concepts:
\cb{Perhaps add limitations?}

\textbf{Items.} \name is a configuration service for existing
data-planes, such as a key-value
store as in our prototype (\cref{sec:implementation}).
We define as \emph{items} the access targets \cey{access targets or targets 
are not used again. can we simply say ``the accessed data''}, \eg key-value 
pairs.
\name manages the
configuration of the data plane, which enables lookup and reconfiguration of
data-plane item replicas\cey{do we need to emphasize replicas here? 
otherwise use``data-plane items''?}.
%Specifically, \name manages replication, placement and migration of the
%configuration: metadata pointers pointing to data-plane items, 
\name interfaces with the existing data store to react to item creation and
migration/reconfiguration by creating and migrating/changing the configuration.

\com{
\gf{An alternative way to phrase the above paragraph is:
"\textbf{Items} are data-plane access targets, e.g., key-value pairs in a data store. The data store replicates items and the coordination service, e.g., \name, manages items: despite items being created, migrated, or reconfigured, the coordination service ensures that clients can look them up."}
}

\com{
\name is a configuration service for data-plane access targets, which we define
as \emph{items}.  \name assumes an existing data plane, such as a distributed
key-value store as we assume in our prototype (\cref{sec:implementation}).
\name manages the metadata regarding the configuration of the data plane,
specifically correct identification of the replicas storing a data item at all
times, even when the item migrates or its replicas change through
reconfiguration.
}

\com{
\textbf{Sites.} We similarly define as \emph{sites}\gf{a bit verbose; start with "\textbf{Sites} are nodes that..."} the nodes that deploy
\name, which we assume to be connected through a network.  For example, sites
can be data centers, the use-case we explore in our prototype and evaluation.
\name can seamlessly make use of sites under the control of different
providers.
}

\textbf{Sites.} Sites are the nodes that deploy \name, which we assume to 
be connected
through a network.  For example, sites can be data centers, which is the 
use-case\cey{Not sure if this is a use case. Instead ``which is the setup''} 
we
explore in our prototype and evaluation.  \name can seamlessly make use of
sites under the control of different providers.

\textbf{Clients requests.} Clients interact with \name by submitting item
lookup requests at a site, and \name's availability guarantees apply once the
client request reaches a site. \name does not
improve last mile availability, \eg if the client cannot reach any site.
We make no assumption about which site
clients choose: Clients can submit lookup requests to any site they wish, \eg
because they change their location, or for other reasons, and the lookup
responses are always correct.  However, \name provides guarantees for the
client w.r.t. the 
location of the client-chosen site. Therefore, 
clients choosing nearby sites based on RTTs would be sensible.

\textbf{Zones, authoritative zone, definitive replicas.} \name provides
availability guarantees for any client looking up an item at the granularity of
\emph{zones}.  \name deploys zones along sites, and a zone encompasses ``the
set of distributed system components, including servers, routers, network
links, etc., that a user depends on for
availability''~\cite{basescu21limix} when looking up items.  The data
plane is zone-agnostic and does not require zone knowledge.  But \name tracks
data-plane item location w.r.t. zones, in order to collocate configuration with
its data. Specifically, \name defines an item's location as the item's
\emph{authoritative zone}: this is the smallest zone containing a quorum of the
item's data plane replicas.  Then, in the same zone, \name maintains the most
recent state of the item's configuration, \ie the \emph{definitive
configuration/metadata replicas} for that item. In contrast, non-authoritative
zones may have only eventually-consistent configuration replicas for those
items.

\textbf{Goals.}
\name has the following objectives:

\begin{itemize}[noitemsep]

\item \textbf{Availability guarantees.} Provide strong availability guarantees
that might be legally or contractually mandated to hold at all times, even during item
migration.

\item \textbf{Simultaneous constraints.} Satisfy simultaneous sovereignty
and locality constraints.
For example, a Germany constraint is more restrictive in placement,
while an EU constraint protects a larger set of users.

\item \textbf{Load balancing.} Spread workload for item lookup
\eg avoid overloading small zones with global accesses.
A user imposes load only on the zones the user's site is in.

\item \textbf{Dynamic data plane.} Enable dynamic data planes
to migrate data routinely,
without restricting them with \eg static partitions of data across regions.

\item \textbf{Strong consistency.} Enforce that item lookup returns
strongly-consistent items, or be unavailable for that item if 
the latest item version is not reachable.

\item \textbf{Autozoning.} Provide an automatic zoning
capability, which enforces locality constraints for all users,
the vast majority of whom just ``want things to work.''
We desire reasonable but fully-automatic risk-limiting policies
requiring no specific understanding of the workload or administrative effort.
%\veg{There is nothing in the evaluation about load balancing and linearizability} 
\end{itemize}

\section{Design of \name}

\com{	This has been said before and says nothing about this section. -baf
\name is the first distributed coordination architecture that enables global
management while also guaranteeing that localized accesses within a region of
interest are insulated from global failures beyond that region.
}
This section outlines \name's design in a step-by-step fashion for clarity.
We first list
\name's challenges, then introduce \name's per-zone configuration service, and
explain how it limits maintenance loads on local zones.
We then address the problem of
satisfying multiple simultaneous exposure-limiting constraints on one item,
ensuring that lookup replies follow data-plane consistency,
 and handling item migration.

\begin{figure*}[!ht]
    \centering
    \begin{subfigure}[t]{.24\textwidth}
        \includegraphics[width=1.05\textwidth]{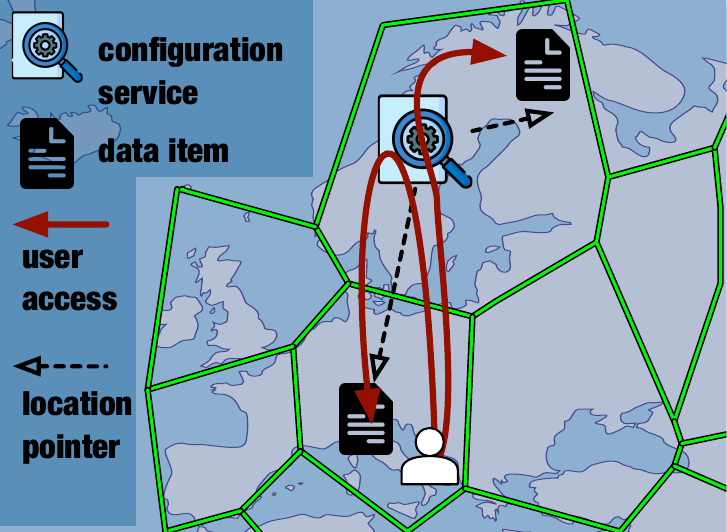}
        \caption{Global CS, holds pointers (ptr.) to all items.}
        \label{fig:strawman0}
    \end{subfigure}\hfill
    %\hspace{-0.2cm}
    \begin{subfigure}[t]{.24\textwidth}
        \includegraphics[width=1.05\textwidth]{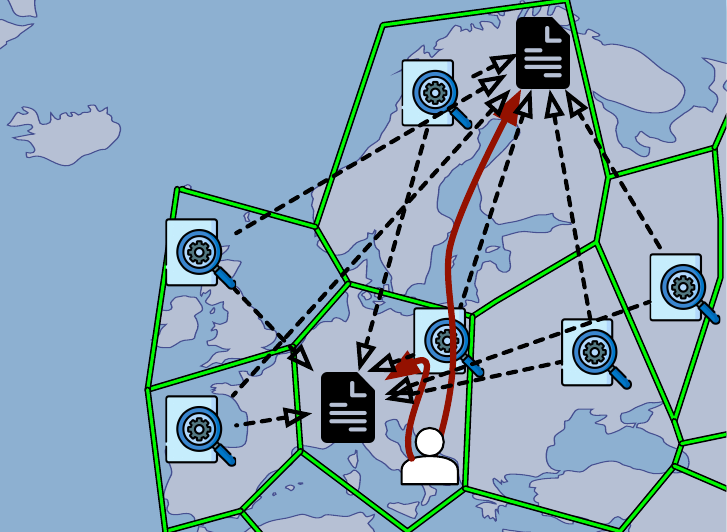}
        \caption{Full CS per zone, each maintaining ptr. to all items.}
        \label{fig:strawman2}
    \end{subfigure}\hfill
    \begin{subfigure}[t]{.24\textwidth}
        \includegraphics[width=1.05\textwidth]{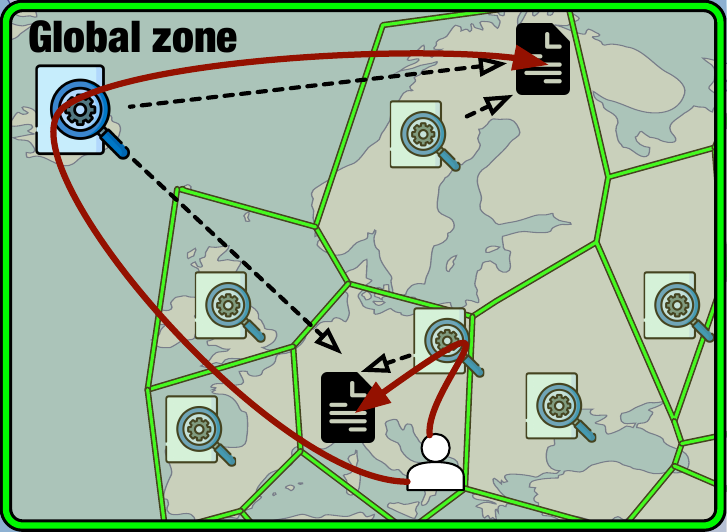}
        \caption{Local CS per zone, each holding ptr. \emph{only} to local items.}
        \label{fig:strawman3}
    \end{subfigure}\hfill
    %\hspace{-0.2cm}
    \begin{subfigure}[t]{.24\textwidth}
        \includegraphics[width=1.05\textwidth]{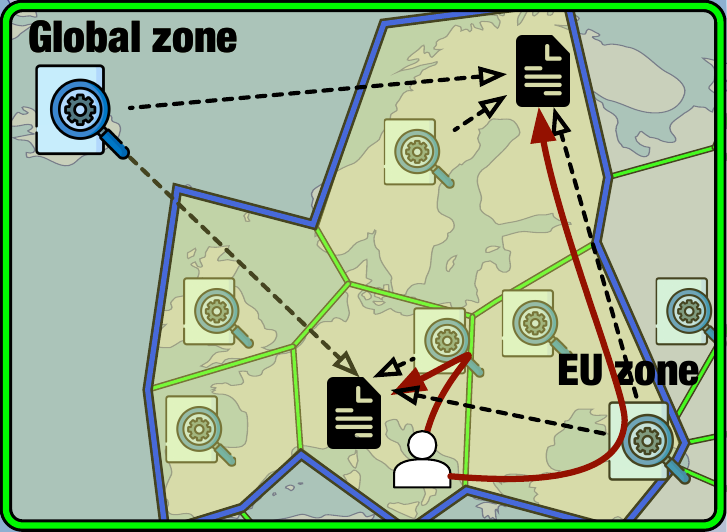}
        \caption{\name CS with overlapping zones, holds ptr. to local items.}
        \label{fig:lookup}
    \end{subfigure}
    \captionsetup{justification=centering}
    \vspace{-0.2cm}
    \caption{Strawmen for the Configuration Service (CS) and their management of pointers (ptr) to data items.}
\end{figure*}

\cb{Change figure to have CS instead of lookup service on left.}

\subsection{Challenges}

\baf{ In the way it is currently written,
	this subsection feels quite overlapping and redundant 
	with the Goals section just shortly before.
}

The goal of local availability under global management,
coupled with the requirement strong consistency for metadata,
imposes key challenges on \name's design.

\textbf{Consistency.} Because data and services must still be
movable and accessible from anywhere under normal conditions,
\name must enable clients anywhere to locate
a data item's current definitive state globally,
while insulating local clients from global failures.
This requirement implies replicating location information
simultaneously across global and local deployment zones,
which in turn exacerbates consistency challenges 
(\cref{sec:lookup,sec:consistency}).

\textbf{Load balancing.} We must ensure that control-plane queries about globally-popular data
do not overload a small, lightly-provisioned zone it may be located in.
\name addresses this challenge by systematically ensuring that
each zone, local or global, serves only clients querying
the service from \emph{within the same zone},
and can therefore spread the access workload
without risking overload from external queries
(\cref{sec:load}).

\textbf{Simultaneous item constraints.}
Data plane items may have to satisfy more than one locality or sovereignty constraint --
such as that local clients in Germany be insulated from failures outside Germany,
\emph{and} that all clients in the EU be insulated from failures outside the EU.
This goal requires that \name allows state replication
across multiple \emph{overlapping} zones
(\cref{sec:simultaneous}).

\textbf{Item migration.}
\name must maintain both strong consistency and local availability
even during object migration:
ensuring, for example, that data migrating from Germany to France
remains immune to failures outside the EU even during the transition.
To address this challenge, \name uses a multi-phase process
to migrate the data's definitive state while maintaining
eventually-consistent location hints in larger zones
beyond the data's origin and destination
(\cref{sec:migration}).

\com{
\textbf{Meaningful availability guarantees.} \name must provide client-centric availability
guarantees meaningful for both the users and the items they access.  \name can
either take zone definitions as input \emph{jurisdictions}, \eg Germany, the
EU, the World, useful when items are constrained by regulatory bounds, for
example. Or, \name can define automated zones -- \emph{autozoning}. Autozoning
is useful for applications where items migrate frequently, without well-knows
predefined patterns. For creating auto-zones, \name borrows the term of Lamport
exposure~\cite{basescu21limix-anon} and round-trip time (RTT) as the exposure
metric. A zone's RTT diameter defines the zone's exposure: a lower RTT diameter
means a lower exposure to remote failures, thus higher availability.
}

\subsection{Item lookup}
\label{sec:lookup}

For zones to act collectively as a unified system, \name needs to enable clients
to find data that can be located anywhere, regardless of the client's zone.  
However, robust
and efficient item lookup is challenging. \cref{fig:strawman0} illustrates
a straightforward but inadequate approach, relying on a central service to
store the configuration for item lookup.  This service increases the client's
exposure beyond the perimeter of the client's and data's common zone.  The single
zone may also become overloaded with requests from all zones.

Of course, this strawman could be easily made scalable by distributing the
configuration service across many/all zones, using standard techniques such as
consistent hashing of keys.
%Physalia~\cite{brooker2020millions}, for example,
%takes a similar approach for its discovery cache.  
However, consistent hashing still
increases a node's Lamport exposure beyond the zone boundaries.
Consider a client requesting data without having the data location cached. To
resolve the location, the client might need to query configuration service
nodes in zones different from the zone holding the requested data.  Partitions
might prevent the client from reading the data location, even though they might
not isolate the client from reaching the data itself.  The problem with this
approach is that data and corresponding metadata are not collocated in the same
zone, and hence lack fate sharing~\cite{clark88design}.

\name thus needs to ensure that a client in a given zone can always find an item
within the same zone using \emph{only} resources within that zone.  Efficiently
collocating data and metadata in the same zone so that they have the same
Lamport exposure represents the first challenge for \name, which we address by
having a distributed configuration service \emph{per zone}.

%\baf{ and at least summarize how that's solved!  i.e., by having a distributed discovery service per zone. }

% there should be a zone for any two users

%\baf{ it feels like we're missing a "Control plane consistency"
%	subsection of some kind, perhaps about here. }

\subsection{Configuration service consistency}
\label{sec:consistency}

As a next strawman approach addressing the challenge above,
we could replicate \emph{all} lookup pointers in \emph{all} zones,           as
depicted in \cref{fig:strawman2}.  However, this strawman introduces a
consistency challenge: Because all zones' configuration services store pointers
to all items, when an item is deleted or migrated, regardless of the zone 
where
the item resided, all configuration services in all zones should be updated with
the new pointer.  If we required strongly consistent state for all zones'
configuration services, we would increase an item's exposure to all zones,
which is undesirable.

\name addresses this exposure challenge as follows. Each zone's configuration
service stores strongly consistent pointers only for the items inside the zone.
If an item's configuration changes, for example, only the configuration service
in the item's zone needs to update the item's definitive configuration (\ie
location metadata) (see \cref{sec:setting}) immediately, on the critical path.
%\baf{Is "definitive" the right term?  Or "primary"?  We need to pick a term and
%use it consistently throughout the paper. }
%\cb{I fixed the issue, is it better?}
Other zones' configuration services may update their metadata lazily
with eventual consistency, outside the critical path of the item
reconfiguration.

With this approach, each zone has its own configuration service that stores
``location hints'' for where an item was last known to be located.  But this
strawman invites a second challenge: How do clients locate items given that
configuration services might (temporarily) store outdated pointers?  We
distinguish two causes of outdated pointers. The first reason is item
migration, when \name updates pointers outside the critical path.  Could a
client be unable to find an otherwise reachable item when the item migrates?
\name addresses the challenge by temporarily storing a \emph{Permanently moved
to} marker at the item's old location.  On encountering this marker, a client
follows it to the new location. \name prevents long indirection paths by
eventually updating all pointers, after which it deletes the marker.  The
client stops following pointers when it reaches the item's
authoritative zone (see \cref{sec:setting}); \name coordination ensures there
is a single authoritative zone per item. \cref{sec:migration} provides a
detailed description.

The second reason for items to be outdated is during partitions. If \name
cannot reach some zones' configuration services, pointers will be stale. The
main challenge is to ensure that clients do not return stale items because of
stale pointers, which would break strong consistency.  \name clients rely on
authoritative zone indicators, as explained above, to decide whether the
pointers point to the most recent item version. However, there is one remaining
issue when partitions heal and several migrations are in place: Pointer updates
might arrive out-of-order, causing an old update to overwrite a newer one.  We
use versioning for pointers to avoid this situation. Every pointer update
increases the pointer version and a pointer update occurs only if the update
has a higher version than the existing pointer. The update makes use of the
compare-and-swap primitive offered in the API of most strongly-consistent KV
stores. \cref{sec:migration} provides a detailed description.
\subsection{Lookup load on (small, local) zones}
\label{sec:load}

The above strawman invites the question: What is the
load on each zone's configuration service?
Consider updating
the configuration service after an item insertion or migration. Either the
destination zone could push the new item location to all zones,
or the client's zone could pull the item location on demand.
Both approaches
incur $O(n)$ load and communication overhead \textit{per client request}
for $n$ zones.

\name instead spreads the lookup loads
and limits query burden on small
zones by organizing a default overlapping global zone.
In our next strawman
illustrated in \cref{fig:strawman3}, local zones store the location only for
their local items and serve lookup requests only from local sites.
In contrast, the global zone serves as the backup
%\veg{What is the load of the global zone?} as the 
%master
reference point, whose globally-distributed configuration service knows any item's
location.
Every zone propagates location updates to the global zone.

A client queries only its own
local configuration service and the global one,
without overloading other small zones.
Thus,
instead of $O(n)$, overhead per update becomes $O(1)$.
The global configuration service must service load from all $O(n)$ zones,
but it has server capacity distributed across all $O(n)$ zones
among which to share that load.
Location updates propagate only eventually, off the critical path, to limit
the source zone's exposure to failures beyond its borders.

\begin{figure}[t]
\begin{algorithm}[H]
  \caption{\name: main structs.\cey{not referenced in text}}
\footnotesize
\begin{algorithmic}[1]

\State \Comment {Every site has a Limix Coordinator, exposing the following API:}
\Struct[SiteCoordinator]
    \State $zone$ \Call{ItemLookup}{$key$}
    \State $value$ \Call{ReadConfig}{$key$}
    \State \Call{WriteConfig}{$key, value$}
    \State \Comment {Updates configuration when item $key$ migrates from $srcZone$ to $dstZone$}
    \State \Call{UpdateConfigOnMigration}{$key, srcZone, dstZone$}
    \State \Comment {Returns all zones that the site is part of}
    \State $zones$ \Call{GetZones}{}
\End

\State \Comment {A Limix Zone stores the configuration and offers the typical API of a strongly consistent KVStore for pointer reads and writes}
\StructExt[ZoneConfigService]{KVStore}
    \State $value$ \Call{ReadPointer}{$key$}
    \State \Comment {Atomically swaps using CAS (compare and swap) $key$'s prior value with $newValue$ if condition evaluates $true$ on the prior value}
    \State $oldValue$ \Call{WritePointerCAS}{$key, newValue, condition$}
    \State \Comment {Returns true if the zone is authoritative for item $key$}
	\Procedure {IsAuthoritative}{$zone, key$}
		\State $zonePointer \gets$ \Call{ReadPointer}{$key$}
		\If{$zonePointer$ == $zone$}
			\Return $True$
		\EndIf
		\Return $False$
	\EndProcedure
\End

\end{algorithmic}
\label{alg:structs}
\end{algorithm}
\vspace{-0.6cm}
\end{figure}

\begin{figure}[t]
\begin{algorithm}[H]
\caption{Item lookup that the client calls at site $site$}
\footnotesize
\begin{algorithmic}[1]

%\Type $KV$

%\EndType

\Procedure{ItemLookup}{$site,key$}
    \For{$zone \in$ \Call{GetZones}{$site$} \textbf{in parallel}}
        \State \Comment{Recursively follow pointers to the authoritative zone}
        \While{$zone \not= nil$ and \Call{IsAuthoritative}{$zone, key$} $\not= true$}
            \State $zonePointer \gets$ \Call{ReadPointer}{$zone,key$}
            \State $zone \gets zonePointer$
        \EndWhile
        \Return $zone$
    \EndFor

\EndProcedure
\end{algorithmic}
\label{alg:item-lookup}
\end{algorithm}
\vspace{-0.6cm}
\end{figure}

%\paragraph{Challenge 3: Item placement.}
\subsection{Item placement and zone overlap}
\label{sec:simultaneous}

\com{
\veg{Does Limix provide a kind of hierarchical failure model? Limix zones can
be described with a tree where larger overlapping zones are parents of
multiple smaller zones. An illustration based on that tree will be useful to
show the reader how items and pointers are placed, how they are migrated and
how data is forced to be unavailable during reconfiguration. I think that may
complete the information given in figure 2. If this new proposed figure is
included in the paper, it may be appropriate to do it in concordance with the
experiments. So it can be used to describe a larger overlapping zone (parent
node in a tree) with a network partition and the items stored at children
nodes that are not affected.  How does reconfiguration affect items'
availability in that tree? Where is the critical path in that tree?}
}

Aside from the global zone, we assumed so far that local zones are disjoint.
This assumption has a significant limitation, however:
it cannot support \emph{simultaneous} exposure-limiting policies,
which may apply by law or contractual obligation.
An item located in Germany may need to be accessible by clients in Germany
with Lamport exposure limited to sites within Germany,
\emph{and} ensure that any client in the EU can access the same item
with exposure limited to the EU.
With the above strawman configuration service, unfortunately,
EU clients outside Germany must query the global configuration service,
yielding global (not EU) Lamport exposure.

\com{   I think this is too subtle and likely confusing here; simplify.
The data plane easily supports
overlapping zones, even for strongly-consistent data.
Using standard
geo-replication, an item is considered within a zone only if a consensus
majority or quorum~\cite{malkhi98byzantine,cachin19asymmetric} (including the
leader replica for leader-based consensus) of the item's replica sites is in
that zone. In our example, a majority of replicas would be in Germany, and the
remaining ones in the EU, ensuring our target Lamport exposure for each zone.
}

\com{
With the above strawman configuration service, unfortunately,
EU users outside Germany must query the global configuration service,
yielding global Lamport exposure.
}

We address this problem
by allowing local zones to overlap.
All zones have a configuration service, and every zone propagates location updates
to larger overlapping zones, up to the global zone. 
The global zone still acts as a master
reference, holding pointers to all items.
Update overhead
increases slightly compared to the single global zone, from $O(1)$ to $O(v)$,
where $v$ is the maximum overlap depth. However, this
approach limits Lamport exposure to smallest zone
containing both the item and the client accessing it.
Location updates propagate only eventually, outside the
critical path, to limit the source zone's exposure to failures outside its
borders. \cref{fig:lookup} illustrates the final configuration service
architecture.

\com{
Because zones store disjoint data slices, zones do not need to coordinate or
depend on each other for the data plane. Unfortunately, this assumption has a
significant limitation: it cannot support \emph{simultaneous} exposure-limiting
policies that may apply, by law or contractual obligation. For example, German
items might follow country-specific data regulation for local users \textit{and}
abide by EU data policies for EU users.
}

%\cb{Should we move the part below to a new subsection?}

\textbf{Item lookup revisited.}
We review item lookup in the final \name configuration service architecture.
We first describe item lookup in a simplified scenario where items are immutable
and do not change location, and the lookup service pointers of all zones are
up-to-date. In this scenario, we do not need to worry about pointer or item
consistency. The next section (\cref{sec:migration}) addresses consistency and migration.

Alg.~\ref{alg:item-lookup} describes item lookup. A client calls the lookup
function on a site's coordinator, passing the item key as a parameter. The
coordinator sends parallel lookup queries to the configuration services of all
site's zones.  The client follows the pointers in the responses until it finds
the authoritative zone, or returns nil.

Why does item search not degrade the exposure
guarantees? Each of the parallel lookup queries accesses the zone-private KV
store, having dependencies only inside the zone. Other parallel searches might
hang in case of partitions or slow performance. However, because each search
executes and completes independently, parallel searches do not affect each
other. Of these zones, the ones that reply first with a pointer chain leading
to the authoritative zone determine the client's exposure for that item. If at
least one such set of zones is partition-free, the client is
\textit{guaranteed} to find the item.  Thus, the coordinator bounds a client's
exposure to the smallest zone of the client that contains the item.

% Below comes the moved parts from the old detailed service

\begin{figure}[!t]
\begin{algorithm}[H]
\caption{Location update during item migration}
\footnotesize
\begin{algorithmic}[1]

\Procedure{UpdatePointerOnMigration}{$key, srcZone, dstZone$}
    \State $oldPointerVersion \gets$ \Call{ReadPointer}{$srcZone, key$}$.Version$
    \State $pointerVersion \gets oldPointerVersion+1$
    \State \Call{VersionedWrite}{$srcZone, key, dstZone, pointerVersion$}
    \State \Call{VersionedWrite}{$dstzone, key, True, pointerVersion$}

    \DoParallel
        \State \Call{UpdatePointerOuterZ}{$srcZone, key, False, pointerVersion, <$}
        \State \Call{UpdatePointerOuterZ}{$dstZone, key, dstZone, pointerVersion, \leq$}
    \End
    \State \Comment {Background garbage collection after all pointers were updated}
    \State \Call{VersionedWrite}{$srcZone, key, nil, pointerVersion$}
\EndProcedure
\end{algorithmic}
\label{alg:lookup-update}
\end{algorithm}

\vspace{-0.6cm}
\end{figure}

\begin{figure}[!t]
\begin{algorithm}[H]
\caption{Helper functions at zone $zone$.}
\footnotesize
\begin{algorithmic}[1]

\Procedure{VersionedWrite}{$zone, key, value, version, comparator$}
    \State \begin{varwidth}[t]{\linewidth}
        \Call{WritePointerCAS}{$zone, key, value||version,$ \par
        \hskip \algorithmicindent $comparator(crtVersion, version)$}
    \end{varwidth}
\EndProcedure

\Procedure{UpdatePointerOuterZ}{$key, value, version, comparator$}
    \For{$outerZone \in$ \Call{getOuterZones}{$zone$} \textbf{in parallel}}
        \State \Call{VersionedWrite}{$outerZone, key, value, version, comparator$}
    \EndFor
\EndProcedure

\end{algorithmic}
\label{alg:zone-helper}
\end{algorithm}

\vspace{-0.6cm}
\end{figure}

\subsection{Lookup during item migration}
\label{sec:migration}

The placement of items may need to change, for example due to client-perceived
performance and load-balancing algorithms. 
Or this could be caused by a policy change,
\eg a new constraint on which existing zone(s) a particular item
is allowed to be placed in or migrated to.  Because items involve
strongly-consistent state, a challenge is that \name must maintain strong consistency
for configuration during item's migration.
Three questions arise at this point: (a) How can we ensure that the coordinator
locates the latest version of an item, even during migration or
partitions? (b) Given that clients could update data plane items anywhere in
the
system, how do we ensure another distant client, located in a different
zone, finds a specific item? (c) How do we ensure item search does not
degrade
the exposure-limiting guarantees?

\com{
We make the previous scenario realistic by allowing item updates (\ie inserts,
writes, deletion) and item migration.
%If the underlying data store updates an item in zone $Z$, it sends a callback
%to the lookup service of zone $Z$.  Similarly, if the data store migrates an
%item from $Z_1 \rightarrow Z_2$, then it notifies the lookup services in those
%zones.
Three questions arise at this point: (a) How can we ensure that a client
locates the latest version of an item, even during migration or
partitions? (b) Given that clients could update data plane items anywhere in
the
system, how do we ensure another distant client, located in a different
zone, finds a specific item? (c) How do we ensure item search does not
degrade
the exposure-limiting guarantees?
}

%The goal of the pointer update is to maintain the
%invariant that each zone's lookup service \textit{eventually} stores pointers
%to all the local zone items. 

Alg.~\ref{alg:lookup-update} depicts item migration from the item's
authoritative zone $srcZone$ to $dstZone$.
When \name receives a callback from the data store that an item
is being migrated, it does the following:
(0) Record the most recent pointer version, taken from the item's authoritative zone, \ie $srcZone$,
and increment it (lines 2-3).
We
use pointer versioning for the pointer updates below to avoid old pointers overwriting newer ones ($versionedWrite$ in Alg.~\ref{alg:zone-helper}).
(1) Update the pointer in the old zone with a forwarding reference, indicating the item
    is being migrated to the new zone and the old zone should no longer be considered authoritative (line 4);
%2. Migrate the item's data-plane state to the new site;
(2) After item migration, update the pointer in the new zone indicating migration is complete
    and the new zone is now authoritative, meaning that item is now usable at its new site (line 5); 
(3) In parallel and outside the critical item lookup path,
update discovery service information in all zones
    for the item's old and new location (lines 6-8); and
(4) Garbage collection: The item's configuration may finally be deleted in the old zone
(line 9).

The algorithm above guarantees that any client can locate an 
item,
even during/after migration.  After step 1, all clients following existing
pointers for that item (using Alg.~\ref{alg:item-lookup}) that lead (directly
or indirectly) to the former authoritative zone learn that the item is being
migrated to the new zone. After step 2, when the item finishes migrating, all
clients following existing pointers locate the new authoritative zone.

Up to now, clients can find the item, but through a potentially longer chain.
For example, if the item moved from Germany to France, clients in the US might
first find the pointer to Germany and then follow it to France.  However,
eventually, when all pointers in step 3 finish updating, all clients in the
system can locate the item by following a single pointer. The exposure during
migration is proportional to the RTT between the client and the two zone
locations of the object. Specifically,
if the item migrates $Z_{item} 
\rightarrow Z_{new}$ then a client's exposure for accessing the item
\emph{during migration} is $Z_{client} \cup Z_{item} \cup Z_{new}$, and
$Z_{client} \cup Z_{new}$ after the migration.
Importantly, because pointer updates happen
independently and in parallel, partitions cannot disrupt a client to locate the
item, as long as both the client and the item share a non-partitioned zone,
which is guaranteed to exist within a small RTT from the client and the item
(\cref{sec:autozoning}).

\section{Control Plane Zoning}
\label{sec:autozoning}

\name provides user-centric availability guarantees meaningful for users with
respect to items they access.  \name can define zones as input
\emph{jurisdictions}, \eg Germany, the EU, the World, which are useful when items are
constrained by regulatory bounds, for example.  However, sometimes
administrative zoning policies may not be explicit.  Or users may want formal
guarantees on availability, alongside the administrative policies.  This
section presents one particular autozoning policy that limits Lamport exposure,
and that can use any distance
metric, such as network latency in our design.  For any user accessing an item
of interest from a distance $\Delta$, autozoning limits the access’s exposure
-- availability and performance -- to a perimeter of at most $O(\log N) \times
\Delta$ around the item.

\begin{algorithm}[!t]                                                           
\caption{Autozoning at site $u$}                                                
\footnotesize                                                                   
\begin{algorithmic}[1]                                                          
\Procedure{BuildAutozones}{$u,Sites,nLevels,RTT$}                               
                                                                                
\For{$v \in Sites$}                                                             
    \State $v.Witnesses \gets CompWitnesses(Sites, nLevels, RTT)$               
    \If{RTT[u][v] < $v.Witnesses[u.Level+1]$}                                   
        \State $u.Cluster \gets u.Cluster \cup v$                               
    \EndIf                                                                      
\EndFor

\For{$v \in u.Cluster$}                                                         
    \For{$radius$ in $i * 2^i$}                                                 
        \If{$RTT[u][v] < radius$}                                               
            \State $u.Zones[radius] \gets u.Zones[radius] \cup v$               
        \EndIf                                                                  
    \EndFor                                                                     
\EndFor                                                                         
                                                                                
\For{$zone \in u.Zones$}                                                        
    \State $StartConfigurationService(zone)$                                    
\EndFor                                                                         
                                                                                
\EndProcedure                                                                   
\end{algorithmic}                                                               
\label{alg:autozoning}                                                          
\end{algorithm}

\begin{figure}[!t]
    \centering
    \begin{subfigure}[t]{0.47\linewidth}
        % \centering
        \includegraphics[trim=30 50 30 20,clip,width=\linewidth]{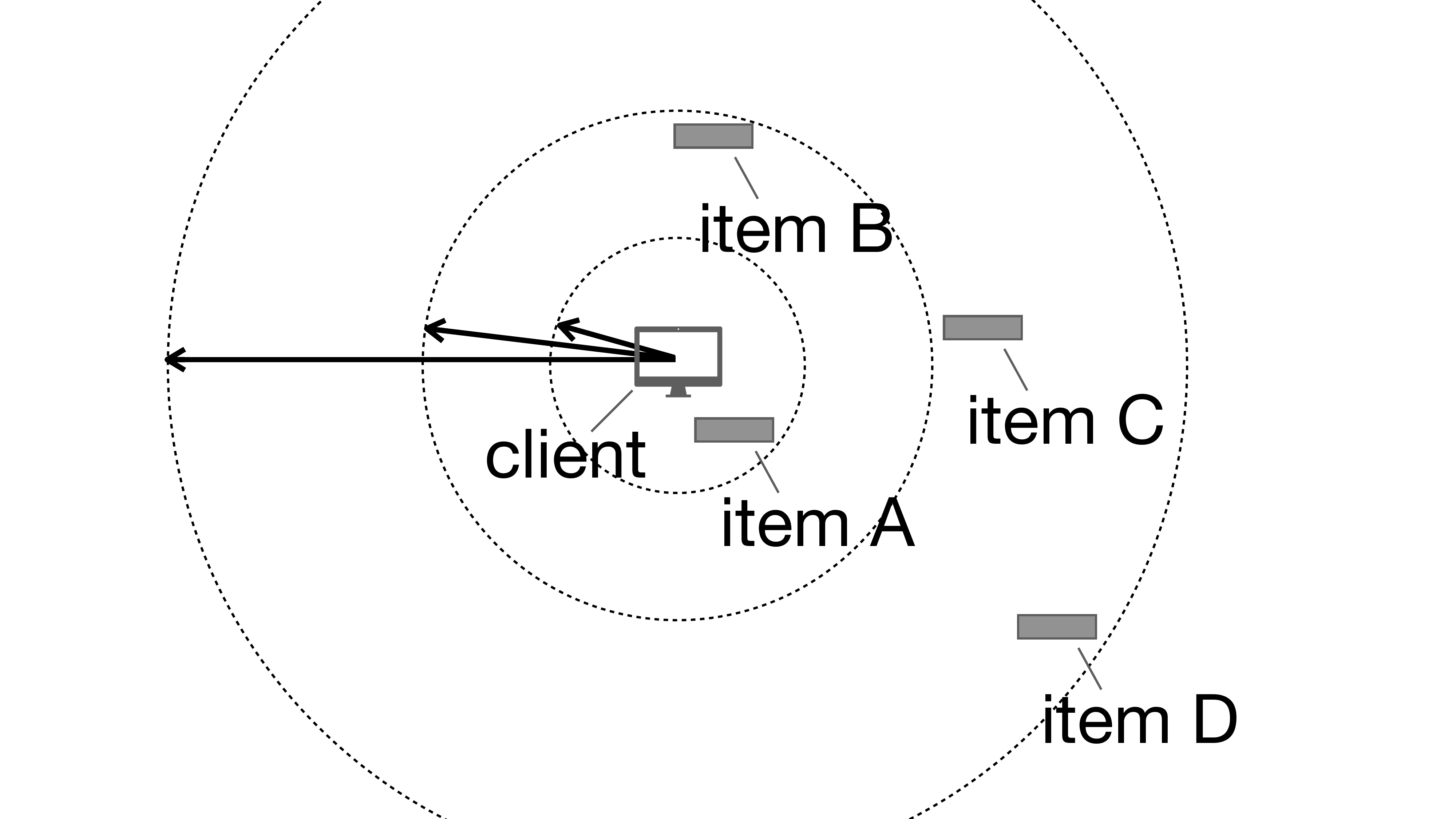} 
        % \caption{Autozoning strawman}
        % \label{fig:concentric_localities}
    \end{subfigure}
    \begin{subfigure}[t]{0.47\linewidth}
        % \centering
        \includegraphics[trim=0 0 0 0,clip,width=\linewidth]{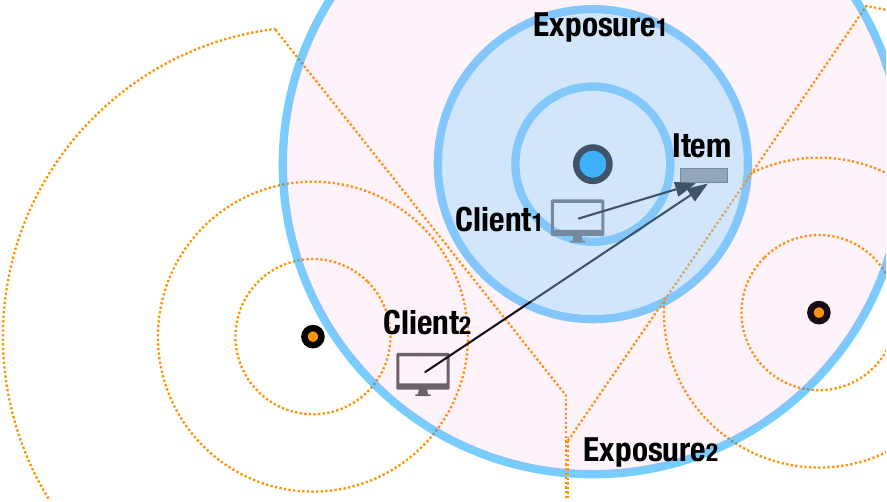} 
        % \caption{Autozoning}
        % \label{fig:overlapping_localities}
    \end{subfigure}
    \caption{Left: Autozoning strawman, bounding exposure for a single client accessing any item. Right: \name autozoning, bounding exposure for \emph{any} client accessing any item.}
	\vspace{-0.4cm}
    \label{fig:concentric_and_overlapping_localities}

\end{figure}

\textbf{RTT as exposure metric.} For creating auto-zones, \name borrows the
term of Lamport exposure~\cite{basescu21limix} and round-trip time (RTT)
as the exposure metric. A zone's RTT diameter defines the zone's exposure: a
lower RTT diameter means a lower exposure to remote failures, thus higher
availability.  During bootstrapping, we build an inter-site RTT map: the sites
measure their pair-wise RTTs, and then each pair of sites averages their link's
RTT value so that the final RTT map is consistent across sites.  Under a fully
connected network, this inter-state RTT map is stable~\cite{haq17cloudpaths},
with prior work~\cite{uluyol20pando} showing less than 6\% month-to-month
difference in median latency on Azure.  Through this initial RTT map, sites
compute the automated zones' membership, at which point bootstrapping
concludes, and \name starts operating without any further assumptions about the
timing or location of partitions.

\com{
\baf{ The text below starts to claim properties very specific to the
	Thorup/Zwick algorithms, but contain no introduction of
	or citations of the relevant Thorup/Zwick papers!  Fix please!  }
}

\textbf{Autozoning strawman.}
To bound the exposure of \emph{a particular client} to any object,
we could simply build overlapping zones centered on the site 
that particular client accesses. 
A na\"ive, non-scalable approach would be to build
many concentric zones of slowly increasing radius.
A more scalable design is to choose exponentially increasing zone
radiuses, as depicted in \cref{fig:concentric_and_overlapping_localities}(a). The
intuition is that that users with a small pairwise RTT are likely
geographically close~\cite{caida-ark}, representing the prioritary localities
for \name. As RTTs increase, localities are less tight. 
The tradeoffs of
exponential zone radiuses are less tight Lamport exposure bounds
for the client accessing any item, with the gain of building a number of sites
logarithmic in network width.

This simple strawman, however, only bounds exposure for clients
accessing the system through the particular site at the center of the zones.
In fact, if a client chooses another site, in line with \name's goals of
supporting dynamic access, then the client does not have any exposure guarantee.
We show next how to build zones that bounds exposure for any client
accessing any object through any site.

\textbf{Compact-graph approximations.}
Autozoning
builds on techniques from
compact graph summarization theory~\cite{thorup01approximate,thorup01compact}
to guarantee exposure for all client and all objects, while optimizing the number
of created zones, hence optimizing the system overhead.
Autozoning has two goals: (1) Bounding
the exposure of a user accessing an item.  (2) Scaling to large deployments by
incurring a logarithmic load on sites.
For the first goal,
recall that the
exposure of a client locating an item is given by the smallest zone containing both the
client and the item.  Our insight is to use compact graph techniques to formally
guarantee an upper bound on the zone RTT diameter, hence on the user's exposure.
Specifically, we want to ensure there exists a
configuration service that has pointers to any item
which is ``close enough'' to any client.
\com{
the sites in this zone run the configuration a lookup service containing the item's
location, and we want to ensure that these sites are close enough to the user
and to the item.  
}%
Compact graph techniques approximate the distance between any
two nodes -- in our case, between a client's site and an item -- to \textit{at
most} $(2 \times k - 1) \times \overline{uv}$, where $\overline{uv}$ is the nodes' exact distance
  and $k$ is a system parameter. They also provide the path between $u$ and $v$
matching this approximate distance. Autozoning groups the sites on the path
between the client's site and the item into a zone.  Intuitively, building such
zones for all user-item pairs ensures that \textit{any} user $u$ looking up
\textit{any} item $i$ is \textit{guaranteed} to find a ``small enough'' common
zone of diameter \textit{at most} $(2 \times k - 1) \times RTT(u,i)$.

The second goal of autozoning is to scale to large deployments.  \name relies
on two techniques for scaling.  The first one comes from compact graph
approximations: To (recursively) compute the paths for the approximate
distances between \textit{any possible} client-site to item-site pair, each
site only needs to know about $O(\log N)$ other sites, where $N$ is the total
number of sites. Even so, if we built all zones incrementally spanning sites on
each path -- and imposed the zone deployment load on the member sites -- the
cost becomes prohibitive because of the large constants in $c * O(\log N)$.
Instead, we use exponentially increasing zone radiuses, as in the strawman.  As
a result, each site participates in and runs a logarithmic number of zones.
\Cref{fig:concentric_and_overlapping_localities}(b) depicts the autozoning design, omitting
the global zone for simplicity.

\textbf{Zone construction.} Alg.~\ref{alg:autozoning} depicts the zone
construction.  Compact-graph approximations use the sites as landmarks for
approximating distances.  Higher-level sites act as global landmarks to
approximate large distances, whereas lower level sites  act as local landmarks.
Each site obtains level $i$ with probability $N^{-i/k}$ ($k$ is the number of
levels). To approximate distances, each sites maintains a set of sites as
contact points, called its \textit{bunch}. A node $u$ explores sites in
ascending distance from itself, and adds a site $v$ in its bunch if $v$'s level
$l_v$ is no smaller than that of any sites explored so far (including $u$). The
sites $v$ closest to $u$ at every level form $u$'s witnesses (line 3).  The
inverse concept of a bunch is a \textit{cluster}.  $v$'s cluster is the set of
sites around $v$, which are ``close enough'' to know about $v$ as a landmark
(lines 4-5). Every site is a landmark and builds zones along its
cluster, using exponentially increasing RTT diameters (lines 6-11). 

\textbf{Lamport exposure bounds.} From the cluster construction, a node at level
$k-1$ has all nodes in its cluster and, thus, creates a Global zone.  Because
node $w$ builds zones on its cluster, and $u$ and $v$ are in its cluster (in
other words, $u$ and $v$ have $w$ in their recursive bunch), $u$ and $v$ are
guaranteed to be in a zone of diameter at most $D = i * 2^i$, where $i$ is the
smallest such that $i * 2^i \geq (2 \times k - 1) \times RTT(u,v)$. By
construction, \textit{any two} sites $u$ and $v$ -- alternatively, a user
contacting site $u$ to look up an item stored on $v$ -- are guaranteed to find
such a zone, providing guaranteed bounds on the Lamport exposure.

\textbf{Load.} The size of a node's bunch is a key property
determining the number of zones that a site is a member of.  From the
probability distribution of level assignment, we expect to accept approximately
$B = \frac{1}{n^{-1/k}}$ nodes into $u$'s bunch at each level $i$. Thus, each
node's bunch has, with high probability, size $|Bunch_u| \approx B \times k
= B \times \log_B(N)$, which upper bounds the number of zones $u$ participates
in. Factoring in the exponential zone diameter increase, $u$
participates on expectation in a polylogarithmic number of zones, or $O(\log
N)$.

\section{Implementation}
\label{sec:implementation}

We implemented a \name prototype of a configuration service interfacing an
existing data store. \name stores its configuration in a per-zone
strongly-consistent KV store. To store the configuration, each zone uses
\crdb~\cite{cockroachdb}, a widely-used strongly-consistent data store.
Although \crdb has rich functionality, \name only uses its basic KV store API
read and compare-and-swap. A \name coordinator runs on every site, providing an
API to query items. Our implementation is written in
Go, with bash scripts for test infrastructure.

\textbf{Startup.} Each site in \name runs a startup script, which takes an
input a list of participating sites and either jurisdictions or autozoning with
the number of levels.  In the case of jurisdictions, the zone membership is
given.  For autozoning, each script measures its RTT to all other sites,
obtains each site's level, and runs locally the Alg.~\ref{alg:autozoning} to
establish the zone membership.  Then the scripts communicate in order to start
a configuration store, \ie \crdb instance, per zone.

\textbf{Processing lookup requests.} Each site runs a \name coordinator that
provides a client API $itemLookup(key): (zone,logical
timestamp)$ to look up strongly-consistent data-plane items. To answer queries,
each site's coordinator queries the configuration stores running on that
site, through site-local configuration store connections using the Go $pq$
driver~\cite{pq} (a PostgreSQL-compatible driver). The configuration stores on
each site communicate with other configuration stores within the zone,
according to \crdb-internal implementation.  A \name coordinator running on a
site part of zone $Z$ receives callbacks from the existing external data store
when the data store creates, updates, deletes or migrates an item in that zone.
For simplicity, we assume an existing external \crdb instance for data items.
Through callbacks, the \name coordinator updates pointers and authoritative
zone information through site-local configuration store queries.

\cb{Add discussion: using a different data store or config store}

%The \name prototype manages configuration at a per-key granularity.  however,
%a future optimization is a coarser grained approach, with one flag per
%multiple (related) keys.

\com{
\paragraph{\name Large-Scale Simulator.} To analyze the latency and availability of \name at larger scales, we created a simulator. The simulator uses the zone creation component and, in a single process, simulates the zone executor. In \crdb, the writer and the reader contact the lease holder replica. Additionally for writes, the lease holder replica contacts a majority of replicas (the fastest to reply). We determine the replicas by hashing a key R times (R is the replication factor) and computing each result modulo the zone membership to determine the R replica nodes. The first replica is also the one that holds the lease. Because zones' membership does not change, the leaseholder nodes do not change either. Similarly, we determine the replicas of the flags by hashing their keys, which we use to simulate flag reads and writes.
}

\section{Evaluation}

\com{
\begin{figure*}[!ht]
    \centering

    \begin{subfigure}[t]{0.30\linewidth}
        % \centering
        \includegraphics[trim=0 0 0 0,clip,width=\linewidth, height=0.60\linewidth]{figures/guarantees-limix.pdf}
        \caption{Limix}
        \label{subfig:limix}
    \end{subfigure}
    \begin{subfigure}[t]{0.30\linewidth}
        % \centering
        \includegraphics[trim=0 0 0 0,clip,width=\linewidth, height=0.60\linewidth]{figures/guarantees-physalia.pdf}
        \caption{Physalia}
        \label{subfig:physalia}
    \end{subfigure}
    \vspace{-0.1cm}
    \caption{Comparison of \name to Physalia for different failure scenarios.
	\baf{ Comparison of *what property*?  (availability?) }
	\baf{ Also, this figure isn't yet referenced or discussed in the text}
    }
    \label{fig:limix_vs_physalia}
    \vspace{-0.2cm}
\end{figure*}
}%

%\com{
We evaluated \name's resilience to network partitions and overheads.
\com{
We evaluated the following aspects of \name: resilience to network partitions, latency of global manageability, and
overheads, and compared \name with cell-based architectures.
}
Our first experimental setup considers jurisdictions, and evaluates \name's overhead.
Our second experimental setup focuses
on \name autozoning and its resilience, and compares \name with Physalia.
%}

\cb{To be moved to appendix:
We also ran an expected overhead analysis on
larger scale networks. 
}

\com{
We experimentally demonstrate \name's theoretical availability guarantees, for
both predefined jurisdictions (\cref{sec:eval:jurisdictions-guarantees}) and
autozoning (\cref{sec:eval:autozoning-guarantees}).  Furthermore, we show that
\name provides its increased availability with minimal resource overheads
in a pay-as-you-go jurisdiction deployment
(\cref{sec:eval:jurisdictions-overheads}) and with logarithmic overheads for
autozoning (\cref{sec:appendix}). Lastly, we show \name's
applicability on real-world setups and scenarios (\cref{sec:eval:real}).  We
run the experiments on both (1) a local cluster with CAIDA-based
geo-distributed RTTs, and (2) an Amazon Web Services (AWS) setup that spans 20
geo-distributed sites.
}

\com{This might go in the appendix}%

\com{
\baf{	Drop this disclaimer, unless something in the NSDI CFP says that
	something like it is needed.
	(Or if it's needed, rephrase to be more specific and believable.) }
This work does not raise any ethical issues.
}%

%\subsection{Experimental setup}
\label{sec:setup}

\textbf{Testbeds.} We used two testbeds for our experiments, both orchestrated using
Kubernetes. The \textbf{cluster testbed} runs 40 Kubernetes site in a local cluster
with a simulated network, which allows for more expeerimental
flexibility. Each site requests 15GiB memory and one
hyperthread of an Intel Xeon Gold 6240
CPU @ 2.60GHz. The delays between sites represent real-world delays of a
globally distributed topology, as follows. Using CAIDA's Archipelago (Ark)
Measurement Infrastructure~\cite{caida-ark}, we first selected 90 monitors of
types infrastructure, research or education (Africa 8, Asia 13, Europe 26, North
America 31, Oceania 4, South America 8) and selected 40 random monitors for our
site locations. To compute the RTT between two monitors at geographical
distance $d$, we averaged the two monitors' median RTT for that distance as
reported by CAIDA. The minimum RTT between two different monitors is 0 and the
maximum is 602.25 ms.

\com{
\gf{
\textbf{Testbeds.}
We used two testbeds for our experiments: (1) a local cluster with a simulated the network, allowing extra experimental flexibility, and (2) a testbed that uses machines from different AWS regions, providing a real-world setup.
In both cases, we use Kubernetes for orchestration.

\textbf{Cluster testbed.} The cluster testbed runs 40 Kubernetes sites in a local cluster
with a simulated network.
Each site requests 15GiB memory and one hyperthread of an Intel Xeon Gold 6240 CPU @ 2.60GHz.
For network delays between sites we use real-world delays of a
globally distributed topology that is based on CAIDA's Archipelago (Ark)
Measurement Infrastructure~\cite{caida-ark}.
Specifically, we first select 90 monitors of
types ``infrastructure'', ``research'', or ``education'' (Africa 8, Asia 13, Europe 26, North
America 31, Oceania 4, South America 8).
Then, we select 40 random monitors for our site locations.
To compute the RTT between two monitors at geographical distance $d$, we average the two monitors' median RTT for that distance as
reported by CAIDA.
RTTs between two different monitors are in the 0 ms to 602.25 ms.

\textbf{Physically geo-distributed testbed.}
For the real-world testbed we run our experiments on AWS, deploying the solutions across 20 sites
}
}

The real-world testbed is deployed on Amazon Web Services (AWS)
and spans 20 sites
(US: East 4, West 4; Canada: central 2, Asia-Pacific: SouthEast 2,
NorthEast 3, EU: Central 1, West 2, South 2).
The server at each site has 16 GiB of DDR4
SDRAM, up to 10 Gbps bandwidth and 2vCPU on 3.1 GHz Intel Xeon processors. The
minimum RTT we measured between sites is 0.43 ms and the maximum is 250.31 ms.

%20 AWS regions with 2 availability zones each, each AZ
%running one r5.large machine in each (US: East 4, West 4; Canada: central 2,
%Asia-Pacific: SouthEast 2, NorthEast 3, EU:
%Central 1, West 2, South 2)
%North 2; South America: East 2).  Each machine has
%16 GiB of DDR4 SDRAM, up to 10 Gbps bandwidth and 2vCPU on either 3.1 GHz Intel
%Xeon or 2.5 GHz AMD processors.  The measured RTT diameter is 414 ms.

\cb{Should we add here a PDF of RTTs between these sites?}

\begin{figure}
\centering
\begin{tikzpicture}

\pgfplotsset{
    legend entry/.initial=,
    every axis plot post/.code={%
        \pgfkeysgetvalue{/pgfplots/legend entry}\tempValue
        \ifx\tempValue\empty
            \pgfkeysalso{/pgfplots/forget plot}%
        \else
            \expandafter\addlegendentry\expandafter{\tempValue}%
        \fi
    },
}

\pgfplotsset{
    % use this `compat' level or higher to use the advanced positioning of
    % the axis labels
    compat=1.3,
}

\usepgfplotslibrary{fillbetween}

\begin{groupplot}[
    group style={
        group name=my fancy plots,
        group size=1 by 1,
        xticklabels at=edge bottom,
        horizontal sep=0pt,
    },
    height=4.5cm,
    ymin=0, ymax=1,
    xlabel style={align=center}, xlabel = {RTT (ms)},
    xlabel style={at={(ticklabel cs:0.75)}},
    xmode=log,
    log basis x=10,
    legend pos=outer north east,
    ylabel style={align=center}, ylabel = {CDF},
    ytick={0, 0.2, 0.4, 0.6, 0.8, 1},
    y axis line style={-},
    xlabel style={at={(ticklabel cs:0.50)}},
    legend style={draw=none},
    legend style={/tikz/every even column/.append style={column sep=0.4cm}},
    grid=major,
    grid style={opacity=0.3},
]

\nextgroupplot[xmin=0,xmax=1000,
               xtick={1, 10, 100, 1000},
               xticklabels={1, 10, 100, 1000},
            %   ytick={-100, -80, -60, -40, -20, 0, 20, 40, 60, 80, 100},
            %   axis y line=right,
            % axis x discontinuity=crunch,
            %   xlabel=\empty,
               width=6.0cm,
            % y axis line style={-},
            % ytick=\empty,
            ]

            \addplot+[line width=1.5pt, mark=none, color=blue, smooth, legend entry={Chen et al.}] coordinates
            {
            (0,  0)
            (1,  0)
            (2,  0)
            (3,  0.016666667)
            (4,  0.038888889)
            (5,  0.072222222)
            (6,  0.105555556)
            (7,  0.15)
            (8,  0.216666667)
            (9,  0.233333333)
            (10,  0.241666667)
            (20,  0.275)
            (30,  0.308333333)
            (40, 0.325)
            (50,  0.5125)
            (60, 0.525)
            (70,  0.583333333)
            (80, 0.7875)
            (90,  0.8)
            (100, 0.8125)
            (110, 0.9125)
            (120, 0.919166667)
            (130, 0.925833334)
            (140, 0.932500001)
            (150, 0.939166668)
            (160, 0.945833335)
            (170, 0.952500002)
            (180, 0.959166669)
            (190, 0.965833336)
            (200, 0.972500003)
            (210, 0.979166667)
            (220, 0.979583334)
            (230, 0.980000001)
            (240, 0.980416668)
            (250, 0.980833335)
            (260, 0.981250002)
            (270, 0.981666669)
            (280, 0.982083336)
            (290, 0.982500003)
            (300, 0.98291667)
            (310, 0.983333333)
            (320, 0.98375)
            (330, 0.984166667)
            (340, 0.984583334)
            (350, 0.985000001)
            (360, 0.985416668)
            (370, 0.985833335)
            (380, 0.986250002)
            (390, 0.986666669)
            (400, 0.987083336)
            (410, 0.9875)
            (420, 0.987916667)
            (430, 0.988333334)
            (440, 0.988750001)
            (450, 0.989166668)
            (460, 0.989583335)
            (470, 0.990000002)
            (480, 0.990416669)
            (490, 0.990833336)
            (500, 0.991250003)
            (510, 0.991666667)
            (520, 0.992083334)
            (530, 0.992500001)
            (540, 0.992916668)
            (550, 0.993333335)
            (560, 0.993750002)
            (570, 0.994166669)
            (580, 0.994583336)
            (590, 0.995000003)
            (600, 0.99541667)
            (610, 0.995833333)
            (620, 0.995972222)
            (630, 0.996111111)
            (640, 0.99625)
            (650, 0.996388889)
            (660, 0.996527778)
            (670, 0.996666667)
            (680, 0.996805556)
            (690, 0.996944445)
            (700, 0.997083334)
            (710, 0.997222223)
            (720, 0.997361112)
            (730, 0.997500001)
            (740, 0.99763889)
            (750, 0.997777779)
            (760, 0.997916668)
            (770, 0.998055557)
            (780, 0.998194446)
            (790, 0.998333335)
            (800, 0.998472224)
            (810, 0.998611113)
            (820, 0.998750002)
            (830, 0.998888891)
            (840, 0.99902778)
            (850, 0.999166669)
            (860, 0.999305558)
            (870, 0.999444447)
            (880, 0.999583336)
            (890, 0.999722225)
            (900, 0.999861114)
            (910, 1)
            (920, 1)
            (930, 1)
            (940, 1)
            (950, 1)
            (960, 1)
            (970, 1)
            (980, 1)
            (990, 1)
            (1000,    1)
            };

            \addplot+[line width=1.5pt, mark=none, color=orange, smooth, legend entry={waikato-1}] coordinates
            {
            (1,   0.00006966213863)
            (2,  0.00006966213863)
            (3,   0.0006966213863)
            (4,  0.00508533612)
            (5,   0.007314524556)
            (6,   0.01330546848)
            (7,   0.02305816789)
            (8,   0.07175200279)
            (9,   0.1194705677)
            (10,  0.1470567746)
            (20,  0.2157436433)
            (30,  0.2546847788)
            (40,  0.2784395681)
            (50,  0.2966213863)
            (60,  0.3203761755)
            (70,  0.3596656217)
            (80,  0.4158133055)
            (90,  0.470846395)
            (100, 0.5212817834)
            (110, 0.5571577847)
            (120, 0.5807035876)
            (130, 0.6042493905)
            (140, 0.6338557994)
            (150, 0.6601880878)
            (160, 0.6870080111)
            (170, 0.7011494253)
            (180, 0.7148728666)
            (190, 0.7334029955)
            (200, 0.7456635319)
            (210, 0.7630094044)
            (220, 0.7876698015)
            (230, 0.8010449321)
            (240, 0.818112156)
            (250, 0.8333681644)
            (260, 0.8422849181)
            (270, 0.8483455242)
            (280, 0.8541274817)
            (290, 0.8599094392)
            (300, 0.8645071404)
            (310, 0.8704980843)
            (320, 0.8768373389)
            (330, 0.8835249042)
            (340, 0.8892371996)
            (350, 0.8953674678)
            (360, 0.901010101)
            (370, 0.9065830721)
            (380, 0.9131313131)
            (390, 0.9180076628)
            (400, 0.9224660397)
            (410, 0.9255311738)
            (420, 0.9284569836)
            (430, 0.9305468478)
            (440, 0.9331243469)
            (450, 0.9361198189)
            (460, 0.9389759666)
            (470, 0.9411354929)
            (480, 0.9430860327)
            (490, 0.945733194)
            (500, 0.948241031)
            (510, 0.9501219087)
            (520, 0.9526994079)
            (530, 0.9546499478)
            (540, 0.9559038662)
            (550 0.957784744)
            (560, 0.9590386625)
            (570, 0.9610588645)
            (580, 0.9619644723)
            (590, 0.9636363636)
            (600, 0.965308255)
            (610, 0.9664925113)
            (620, 0.967816092)
            (630, 0.9687216998)
            (640, 0.9699059561)
            (650, 0.9710205503)
            (660, 0.9721351445)
            (670, 0.9731104145)
            (680, 0.9740856844)
            (690, 0.9754092651)
            (700, 0.9764541971)
            (710, 0.9777777778)
            (720, 0.9788227099)
            (730, 0.9799373041)
            (740, 0.9807732497)
            (750, 0.9817485197)
            (760, 0.9825844653)
            (770, 0.9830721003)
            (780, 0.983420411)
            (790, 0.9840473703)
            (800, 0.9848136538)
            (810, 0.9853709509)
            (820, 0.9857889237)
            (830, 0.9866248694)
            (840, 0.9878091257)
            (850, 0.9885057471)
            (860, 0.9892720307)
            (870, 0.9901079763)
            (880, 0.9908045977)
            (890, 0.9919888541)
            (900, 0.9928247997)
            (910, 0.9933820968)
            (920, 0.9941483804)
            (930, 0.994984326)
            (940, 0.9955416231)
            (950, 0.9965168931)
            (960, 0.9976314873)
            (970, 0.9981191223)
            (980, 0.998467433)
            (990, 0.9993033786)
            (1000,    1)
            };

            \addplot+[line width=1.5pt, mark=none, color=teal, smooth, legend entry={waikato-2}] coordinates
            {
            (1,   0.0006269101168)
            (2,   0.0006269101168)
            (3 ,  0.000783637646)
            (4,   0.000783637646)
            (5,   0.001802366586)
            (6,   0.01034401693)
            (7,  0.01998275997)
            (8,   0.07013556931)
            (9,   0.1013243476)
            (10,  0.1104929081)
            (20,  0.1522607946)
            (30,  0.1685604576)
            (40,  0.1753781052)
            (50,  0.1838413917)
            (60,  0.1955175927)
            (70,  0.2244338218)
            (80,  0.2622835201)
            (90,  0.2949612099)
            (100, 0.3296763577)
            (110, 0.3473082047)
            (120, 0.36219732)
            (130, 0.3747355223)
            (140, 0.393151007)
            (150, 0.4255936055)
            (160, 0.4362510775)
            (170, 0.4471436408)
            (180, 0.456390565)
            (190, 0.4687720398)
            (200, 0.4791160567)
            (210, 0.5743280307)
            (220, 0.5893738735)
            (230, 0.5970535225)
            (240, 0.6131180942)
            (250, 0.6320821252)
            (260, 0.6410155944)
            (270, 0.6486952433)
            (280, 0.6544157981)
            (290, 0.6610767181)
            (300, 0.6678160019)
            (310, 0.6763576522)
            (320, 0.6861531228)
            (330, 0.6956351383)
            (340, 0.7030796959)
            (350, 0.7109160724)
            (360, 0.7184389938)
            (370, 0.7274508267)
            (380, 0.7363842959)
            (390, 0.7477470418)
            (400, 0.7580910587)
            (410, 0.7684350756)
            (420, 0.779954549)
            (430, 0.7898283834)
            (440, 0.8013478568)
            (450, 0.8113000549)
            (460, 0.8231329833)
            (470, 0.8367682783)
            (480, 0.8451532012)
            (490, 0.8547919442)
            (500, 0.8625499569)
            (510, 0.8709348797)
            (520, 0.8789279837)
            (530, 0.88613745)
            (540, 0.8912310947)
            (550, 0.8949925554)
            (560, 0.89945929)
            (570, 0.904474571)
            (580, 0.908784578)
            (590, 0.9123893112)
            (600, 0.9160724081)
            (610, 0.9199905963)
            (620, 0.9232818745)
            (630, 0.9279053366)
            (640, 0.9307264321)
            (650, 0.933077345)
            (660, 0.9369171695)
            (670, 0.9408353577)
            (680, 0.9445184547)
            (690, 0.947417914)
            (700, 0.9505524645)
            (710, 0.9529817412)
            (720, 0.9559595643)
            (730, 0.9579970222)
            (740, 0.9602695714)
            (750, 0.9621503017)
            (760, 0.9641093958)
            (770, 0.9656766711)
            (780, 0.9668521276)
            (790, 0.9687328579)
            (800, 0.9699083144)
            (810, 0.9712404984)
            (820, 0.9731212287)
            (830, 0.9744534127)
            (840, 0.9771177807)
            (850, 0.9795470574)
            (860, 0.9808008777)
            (870, 0.982368153)
            (880, 0.9843272471)
            (890, 0.9857377948)
            (900, 0.9873050701)
            (910, 0.9888723454)
            (920, 0.9900478019)
            (930, 0.9913016221)
            (940, 0.9924770786)
            (950, 0.9935741713)
            (960, 0.9947496278)
            (970, 0.996003448)
            (980, 0.9975707233)
            (990, 0.9985110885)
            (1000,    1)
            };

            \addplot+[line width=1.5pt, mark=none, color=yellow, smooth, legend entry={waikato-3}] coordinates
            {
            (1,   0.0001461560947)
            (2,   0.0001461560947)
            (3,   0.0002192341421)
            (4,   0.004896229173)
            (5,   0.007307804735)
            (6,   0.01731949722)
            (7,   0.02828120433)
            (8,   0.05415083309)
            (9,   0.07717041801)
            (10,  0.092589886)
            (20,  0.1387752119)
            (30,  0.1679333528)
            (40,  0.1818181818)
            (50,  0.1967991815)
            (60,  0.2218649518)
            (70,  0.2538731365)
            (80,  0.2898275358)
            (90,  0.3314820228)
            (100, 0.3643671441)
            (110, 0.3870213388)
            (120, 0.402733119)
            (130, 0.4294066063)
            (140, 0.4571031862)
            (150, 0.4762496346)
            (160, 0.4926921953)
            (170, 0.5048231511)
            (180, 0.5177579655)
            (190, 0.5301081555)
            (200, 0.5513738673)
            (210, 0.5754896229)
            (220, 0.5878398129)
            (230, 0.6034785151)
            (240, 0.6151710026)
            (250, 0.6338789828)
            (260, 0.6412598655)
            (270, 0.6480561239)
            (280, 0.6580678164)
            (290, 0.6658140894)
            (300, 0.6834258989)
            (310, 0.6985530547)
            (320, 0.7065185618)
            (330, 0.7152148495)
            (340, 0.7304881614)
            (350, 0.7447383806)
            (360, 0.7624963461)
            (370, 0.7816427945)
            (380, 0.7965507162)
            (390, 0.8085355159)
            (400, 0.8234434376)
            (410, 0.8327974277)
            (420, 0.8408360129)
            (430, 0.8515054078)
            (440, 0.8583747442)
            (450, 0.8650248465)
            (460, 0.8705057001)
            (470, 0.8751826951)
            (480, 0.8794942999)
            (490, 0.8844636071)
            (500, 0.8885559778)
            (510, 0.891917568)
            (520, 0.8954253142)
            (530, 0.8987869044)
            (540, 0.9014177141)
            (550, 0.9066793335)
            (560, 0.9106986261)
            (570, 0.9149371529)
            (580, 0.9189564455)
            (590, 0.9220257235)
            (600, 0.9250950015)
            (610, 0.9288219819)
            (620, 0.9307950892)
            (630, 0.9332066647)
            (640, 0.9396375329)
            (650, 0.9504530839)
            (660, 0.9539608302)
            (670, 0.9569570301)
            (680, 0.9587839813)
            (690, 0.961414791)
            (700, 0.9631686641)
            (710, 0.9646302251)
            (720, 0.9666764104)
            (730, 0.9681379714)
            (740, 0.9695264543)
            (750, 0.9710610932)
            (760, 0.9724495761)
            (770, 0.9731803566)
            (780, 0.975372698)
            (790, 0.9771265712)
            (800, 0.978368898)
            (810, 0.9792458346)
            (820, 0.9806343175)
            (830, 0.9817304882)
            (840, 0.9832651272)
            (850, 0.984580532)
            (860, 0.9854574686)
            (870, 0.9866267173)
            (880, 0.9878690441)
            (890, 0.9890382929)
            (900, 0.9896959953)
            (910, 0.9908652441)
            (920, 0.9918883367)
            (930, 0.9929845075)
            (940, 0.9942999123)
            (950, 0.995396083)
            (960, 0.9961999415)
            (970, 0.9973691903)
            (980, 0.9983192049)
            (990, 0.9989769073)
            (1000,    1)
            };

            \addplot+[line width=1.5pt, mark=none, color=purple, smooth, legend entry={waikato-4}] coordinates
            {
            (1,   0)
            (2,   0)
            (3,   0.0001193175039)
            (4,   0.0005369287675)
            (5,   0.00155112755)
            (6,   0.00954540031)
            (7,   0.01771864933)
            (8,   0.0507695979)
            (9,   0.07916716382)
            (10,  0.09318697053)
            (20,  0.1368571769)
            (30,  0.1648967904)
            (40,  0.1811239709)
            (50,  0.1944278726)
            (60,  0.2111919819)
            (70, 0.2464503043)
            (80,  0.2876745018)
            (90, 0.3281827944)
            (100, 0.3603985205)
            (110, 0.3821143062)
            (120, 0.399236368)
            (130, 0.4156425248)
            (140, 0.4378952392)
            (150, 0.4580002386)
            (160, 0.4852642883)
            (170, 0.5044744064)
            (180, 0.5171220618)
            (190, 0.5303663047)
            (200, 0.5437895239)
            (210, 0.5722467486)
            (220, 0.6066101897)
            (230, 0.6220021477)
            (240, 0.6366782007)
            (250, 0.6489679036)
            (260, 0.6601240902)
            (270, 0.6677007517)
            (280, 0.675575707)
            (290, 0.6829733922)
            (300, 0.69168357)
            (310, 0.7050471304)
            (320, 0.7188879609)
            (330, 0.7396492065)
            (340, 0.7546235533)
            (350, 0.7678677962)
            (360, 0.7786660303)
            (370, 0.7875551843)
            (380, 0.7971005847)
            (390, 0.8070635962)
            (400, 0.8135067414)
            (410, 0.8206061329)
            (420, 0.8259157618)
            (430, 0.8413673786)
            (440, 0.8494809689)
            (450, 0.8546712803)
            (460, 0.8592650042)
            (470, 0.8649325856)
            (480, 0.8707791433)
            (490, 0.8748955972)
            (500, 0.8804438611)
            (510, 0.8858131488)
            (520, 0.8905858489)
            (530, 0.8968500179)
            (540, 0.9010261305)
            (550, 0.9052022432)
            (560, 0.9104522133)
            (570, 0.9161794535)
            (580, 0.9212504474)
            (590, 0.9250686076)
            (600, 0.9282305214)
            (610, 0.9325259516)
            (620, 0.9354492304)
            (630, 0.9388497793)
            (640, 0.9420713519)
            (650, 0.9448753132)
            (660, 0.9479775683)
            (670, 0.9501849421)
            (680, 0.9521536809)
            (690, 0.9548383248)
            (700, 0.9579405799)
            (710, 0.9602076125)
            (720, 0.96223601)
            (730, 0.9642047488)
            (740, 0.9655769001)
            (750, 0.9672473452)
            (760, 0.968500179)
            (770, 0.9702302828)
            (780, 0.9719007278)
            (790, 0.9741081017)
            (800, 0.975718888)
            (810, 0.976912063)
            (820, 0.9781648968)
            (830, 0.9798353418)
            (840, 0.9811478344)
            (850, 0.9829375969)
            (860, 0.9840711132)
            (870, 0.9852046295)
            (880, 0.9867557571)
            (890, 0.9875909796)
            (900, 0.9892017659)
            (910, 0.9910511872)
            (920, 0.9918864097)
            (930, 0.9933182198)
            (940, 0.9943920773)
            (950, 0.9953466173)
            (960, 0.9965397924)
            (970, 0.9975539912)
            (980, 0.9983295549)
            (990, 0.9991051187)
            (1000,    1)
            };

\end{groupplot}

\end{tikzpicture}

\vspace{-0.5cm}
\caption{
CDF of reconfiguration RTTs.
}
\label{fig:man_workloads}
\vspace{-0.5cm}
\end{figure}

\begin{figure*}[!ht]
    \centering

    \begin{subfigure}[t]{0.23\linewidth}
        % \centering
        \includegraphics[trim=0 0 0 0,clip,width=\linewidth, height=0.70\linewidth]{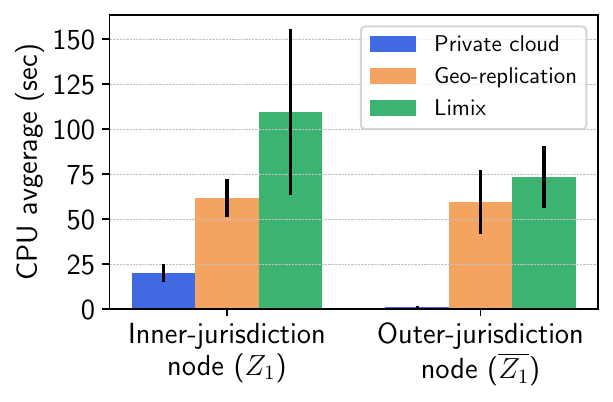}
        \caption{CPU}
        \label{subfig:cpu}
    \end{subfigure}
    \begin{subfigure}[t]{0.23\linewidth}
        % \centering
        \includegraphics[trim=0 0 0 0,clip,width=\linewidth, height=0.70\linewidth]{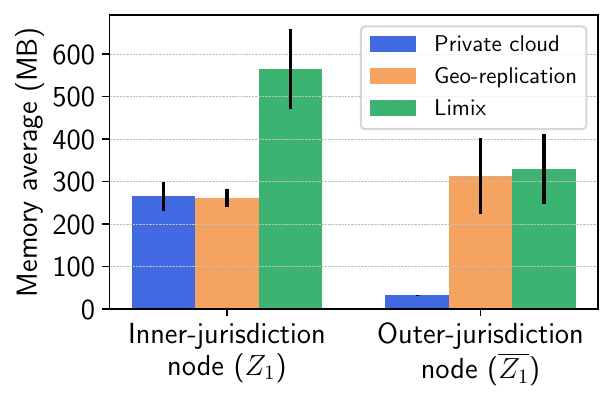}
        \caption{Memory}
        \label{fig:jr-mem}
    \end{subfigure}
    \begin{subfigure}[t]{0.23\linewidth}
        % \centering
        \includegraphics[trim=0 0 0 0,clip,width=\linewidth, height=0.70\linewidth]{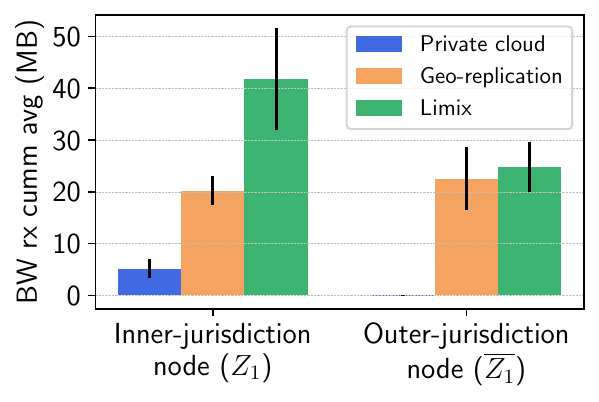}
        \caption{Bandwidth RX}
        \label{subfig:bw_rx}
    \end{subfigure}
    \begin{subfigure}[t]{0.23\linewidth}
        % \centering
        \includegraphics[trim=0 0 0 0,clip,width=\linewidth, height=0.70\linewidth]{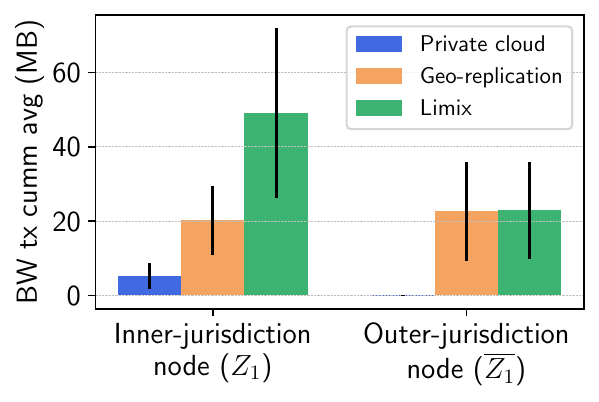}
        \caption{Bandwidth TX}
        \label{subfig:bw_tx}
    \end{subfigure}
    \vspace{-0.1cm}
    \caption{Jurisdictions: compute, memory, and bandwidth overhead.
    %\baf{The first two bars in each graph should be swapped,
    %   so that "Private cloud" is always the first bar
    %   rather than the middle bar.
    %   This would be consistent with Figure 1
    %   and with the fact that "private cloud"
    %   is fundamentally the simplest comparison baseline.}
    }
    \label{fig:jurisdictions_overhead}
    \vspace{-0.2cm}
\end{figure*}

\textbf{Software setup.}
\name autozoning experiments use two levels
for building zones through compact graph approximations,
and zone diameters of $2(i-2)\sqrt{2}^i$, where $i \geq 3$.
For our Physalia implementation, we use cells of up to 50 ms maximum RTT,
based on the 99th percentile write latency reported in the published paper.
\name zones and Physalia cells all run Core \crdb ver.20.1.

\com{
Our baseline experiments run Core \crdb ver.19.2.  Core \crdb manages its
configuration in the so-called \textit{ranges}, for which we keep the default
replication factor of five.  None of \crdb's multi-region patterns described in
the documentation~\cite{crdb-doc1,crdb-doc2,crdb-doc3} seem to                         limit exposure during
reconfiguration, which is what our experiments test. The documentation states
that all configuration ranges \say{\textit{must retain a majority       of
replicas for the cluster as a whole to remain available}}~\cite{crdb-doc3}.
\baf{ drop the gratuitous italics; the quotation marks are enough. }
Unless otherwise specified, \name runs in autozoning mode, and uses 3 levels
for building zones through compact graph approximations.
Please refer to our discussion (Sec~\ref{sec:discussion}).
}%

\com{
\veg{CITE - and preferable archive the page in webarchive}. For the \textit{meta range} that contains the location of cluster
data: \say{\textit{If your cluster is running in multiple datacenters, it's a best
practice to [...] have a copy in each datacenter.}}. For the liveness and system
ranges, the documentation recommends keeping the default of 5 replicas, which
we do.
At the same time, the
documentation states that all these ranges \say{\textit{must retain a majority
of replicas for the cluster as a whole to remain available}}.
Conversely, if a small local zone is disconnected from the rest of the network,
it is unlikely the users in that
zone can reach a majority of range replicas, and their reconfiguration
operations fail.
}

\textbf{Workloads.}
We evaluate \name using configuration writes (W), because they are the
critical operations during gray failures (\cref{sec:motivation}).
Specifically, we use pairs of write
(W-W) operations between pairs of sites.  Each W-W pair concerns the
same item. The operation pair emulates a reconfiguration for that item,
for example after an item migration.
The reconfiguration write issued by the second node has a strong
consistency dependency on the first.  We use the  term ``reconfiguration RTT''
to refer to the RTT between the two writers.  The goal is to understand
the dependencies of the reconfiguration operation to succeed, given a known
prior location of the configuration.
In our experiments, we run the workload as Poisson arrivals, at a rate
of 20 pairs per second.

We use real-world distributions for the reconfiguration RTTs based on
metropolitan traffic traces.  We chose metropolitan traffic traces as they
capture more local traffic, as opposed to backbone links that might miss local
traffic.  Chen et al.~\cite{chen20measuringtcprtt} provide an
RTT distribution over a 10Gbps metropolitan link -- henceforth called trace 1.
We compared this trace to four traces of the Waikato
dataset~\cite{waikatotraces}, whose RTT distribution we extracted using a
similar methodology as~\cite{chen20measuringtcprtt}: we matched data packets
with the respective ACKs. These traces rely on public datasets and the
extraction methodology avoids raising ethical issues.
\Cref{fig:man_workloads} shows the cumulative distribution function (CDF)
for the reconfiguration RTT. Given the similarity of the distributions, and the
fact that trace 1 is more recent, we used trace 1 for generating the
reconfiguration RTTs.

\subsection{Jurisdictions: availability and costs}
\label{sec:eval:jurisdictions-guarantees}

Our first experiment focuses on a simple \name deployment
in which each item exists in only one local zone,
in addition to the default global zone.
In this scenario we consider each of the disjoint local zones
to represent an administratively-defined \emph{jurisdiction},
such as a country.
The experiment
answers the question: \say{\textit{What are the availability benefits and cost
per \name zone in this scenario?}}

\methodology{}
We ran the experiment on the cluster testbed. We focus on one particular jurisdiction
centered around a CAIDA monitor in Europe West,
and choose an RTT radius of 50 ms around
the center, which roughly corresponds to the EU-West jurisdiction. Prior
work has also shown that latencies of roughly 30ms correspond to country-wide
RTTs in Europe~\cite{chen20measuringtcprtt}.

\com{
\baf{ Note that I rephrased the above text so that
	the experimental *focus* is on only one local jurisdiction/zone,
	but it does not say that there *is* (necessarily)
	only one local jurisdiction/zone.
	So adding the $n \times$ height bars to the graphs as discussed
	should now be consistent with the scenario at least, if feasible. }
}

We are interested in the overheads during no-partition conditions, and in the
availability during partitions. For this experiment, we run a synthetic
workload with 2000 pairs of W-W operations: 1000 W-W pairs are for
reconfiguration in EU-West, and 1000 for global reconfiguration. For the
availability experiment, we run the same operations, but we disconnect EU-West
from Global. We denote EU-West by $Z_1$ and Global $\setminus$ EU-West by
$\overline{Z_1}$. In \name, nodes in $Z_1$ are part of two zones, whereas nodes
in $\overline{Z_1}$ are part of one zone.

%\begin{figure*}[!t]
    %\centering
    %\begin{subfigure}[t]{.32\textwidth}
        %\includegraphics[width=1.05\textwidth]{figures/az_1r.pdf}
        %\caption{$R_G=R_C$: Collab. close to partitions.}
        %\label{fig:availability-close}
    %\end{subfigure}\hfill
    %%\hspace{-0.2cm}
    %\begin{subfigure}[t]{.32\textwidth}
        %\includegraphics[width=1.05\textwidth]{figures/az_2r.pdf}
        %\caption{$R_G=2*R_C$: Collab. near partitions.}
        %\label{fig:availability-near}
    %\end{subfigure}\hfill
    %\begin{subfigure}[t]{.32\textwidth}
        %\includegraphics[width=1.05\textwidth]{figures/az_5r.pdf}
        %\caption{$R_G=5*R_C$: Collab. far from partitions.}
        %\label{fig:availability-far}
    %\end{subfigure}\hfill
    %\captionsetup{justification=centering}
   %\vspace{-0.2cm}
    %\caption{Measured availability depending on the proximity of collaborators to partitions.}
%\vspace{-0.4cm}
%\end{figure*}

\begin{figure*}[!ht]
    \centering

    \begin{subfigure}[t]{0.037\linewidth}
        % \centering
        \includegraphics[trim=0 0 298 15,clip,width=\linewidth, height=4.25\linewidth]{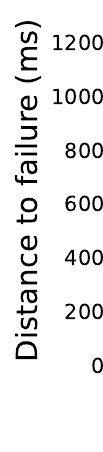}
    \end{subfigure}
     \begin{subfigure}[t]{0.45\linewidth}
        \begin{subfigure}[t]{0.49\linewidth}
            % \centering
            \includegraphics[trim=52 0 0 18,clip,width=\linewidth, height=0.70\linewidth]{figures/guarantees-limix-uniform+princeton-runonAWSNo.pdf}
            \caption{\name on cluster testbed}
            \label{fig:local_limix}
        \end{subfigure}
        \begin{subfigure}[t]{0.49\linewidth}
            % \centering
            \includegraphics[trim=52 0 0 18,clip,width=\linewidth, height=0.70\linewidth]{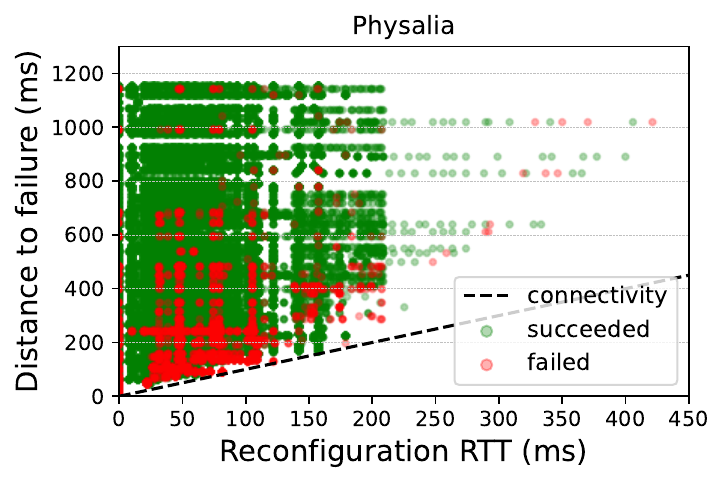}
            \caption{Physalia on cluster testbed}
            \label{fig:local_physalia}
        \end{subfigure}
        % \caption{Local Kubernetes testbed}
        % \label{fig:guarantees:local}
    \end{subfigure}
    \begin{subfigure}[t]{0.037\linewidth}
    % \centering
    \includegraphics[trim=0 0 289 15,clip,width=\linewidth, height=4.25\linewidth]{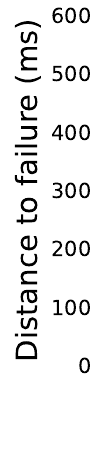}
\end{subfigure}
    \begin{subfigure}[t]{0.45\linewidth}
        \begin{subfigure}[t]{0.49\linewidth}
            % \centering
            \includegraphics[trim=48 0 0 18,clip,width=\linewidth, height=0.70\linewidth]{figures/guarantees-limix-princeton-runonAWSYes.pdf}
            \caption{\name on AWS}
            \label{fig:aws_limix}
        \end{subfigure}
        \begin{subfigure}[t]{0.49\linewidth}
            % \centering
            \includegraphics[trim=48 0 0 18,clip,width=\linewidth, height=0.70\linewidth]{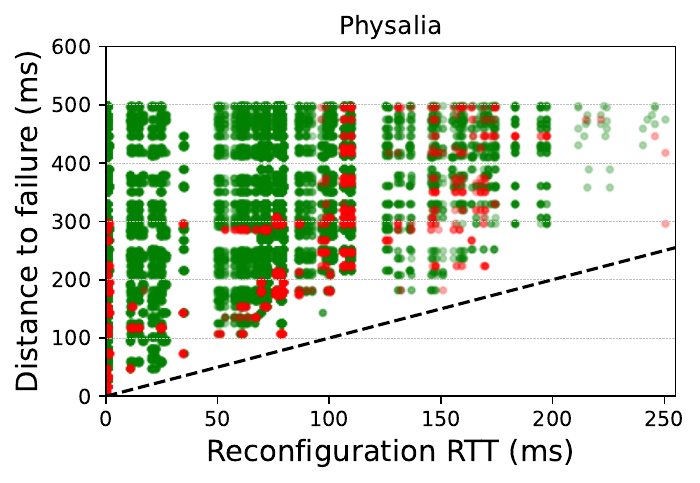}
            \caption{Physalia on AWS}
            \label{fig:aws_physalia}
        \end{subfigure}
        % \caption{AWS testbed}
        % \label{fig:guarantees:aws}
    \end{subfigure}
    \vspace{-0.1cm}
    \caption{Comparison of availability of \name and Physalia for different failure scenarios and testbeds.}
    \label{fig:guarantees}
    \vspace{-0.2cm}
\end{figure*}

\begin{figure}[!t]
    \centering
    \begin{subfigure}[t]{\columnwidth}
        \centering
        \includegraphics[trim=0 0 0 0,clip,height=0.48\linewidth]{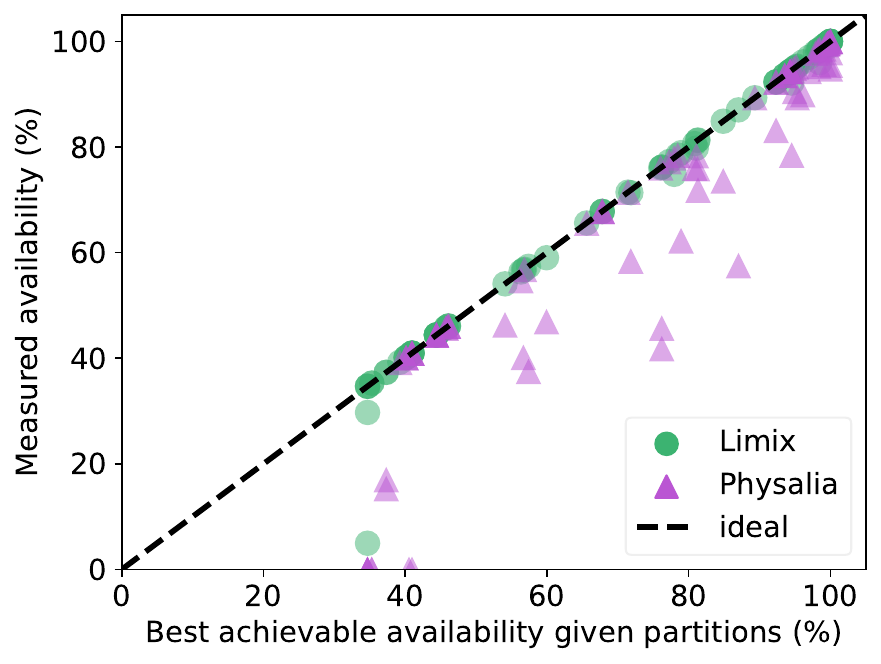}
    \end{subfigure}
    \vspace{-0.5cm}
    \caption{Comparison of \name to Physalia on AWS and realistic workloads.}
    \label{fig:best_vs_actual_availability_AWS}
    \vspace{-0.4cm}
\end{figure}

\textbf{Availability.} We compare \name against two baselines: core \crdb with
geo-replication in the Global zone -- henceforth called Geo, and core \crdb
deployed as a private cloud in EU-West only -- henceforth called PCloud.
Table~\ref{tab:jurisdiction} summarizes the features of the three designs.
Geo offers global configuration management, but reconfiguration in $Z_1$ fails
under partitions. The reason is that, under partition, $Z_1$ cannot reach a
majority of configuration replicas. PCloud succeeds reconfigurations in $Z_1$
under partition, but offers no global configuration management, because all
configuration is in $Z_1$.
The experiment confirms that
\name offers both availability in $Z_1$ \emph{and}
global configuration management.

\textbf{Overheads.}
\label{sec:eval:jurisdictions-overheads}
For the same workload, we report the memory, CPU and bandwidth overheads under
no-partition conditions. \cref{fig:jurisdictions_overhead} shows these
overheads for nodes in $Z_1$ and $\overline{Z_1}$.  As expected, PCloud nodes
have the lowest overheads, but PCloud lacks global manageability.  PCloud
nodes $\overline{Z_1}$
simply
forward client requests to the closest node in $Z_1$,
hence use almost no resources.
The nodes in $Z_1$
running PCloud have lower overheads than Geo nodes because the PCloud
deployment is smaller, and requires  fewer resources for coordination between
nodes, for example.  For its improved resilience guarantees, \name sites in
$Z_1$ spend about 2x the memory, CPU and bandwidth compared to Geo.  The memory
overhead stems from nodes in $Z_1$ running two instances of \crdb: one
corresponding to the inner- and the other to the outer-jurisdiction.  To
explain the CPU and bandwidth overheads, recall that, when a \name site in
$Z_1$ executes a write, it also writes to $\overline{Z_1}$.  However, \name
sites in $\overline{Z_1}$, which do not have an improved resilience guarantee compared
to Geo, have overheads very similar to Geo. We conclude that \name is suitable for
a highly configurable pay-as-you-go deployment, where every extra resource spent provides an immediate
increase in availability.

\begin{table}[t!]
\setlength{\tabcolsep}{2pt}
\small
%\centering
\begin{tabular}{l c c c c}
 \hline
 \multicolumn{2}{l}{Feature} & \multicolumn{2}{c}{\crdb}  & \name \\ %[0.5ex]
 \multicolumn{2}{l}{ } &  Geo-replication & Private cloud & \\ %[0.5ex]
 \hline\hline
Availability $Z_1$ reconfig. & &0\% & 100\% & 100\% \\
Global configuration mgmt. & &\ding{51} & \ding{53} & \ding{51} \\
 \hline
\end{tabular}
\caption{Jurisdictions: features, measured availability.}
%\vspace{-0.5cm}
\label{tab:jurisdiction}
\vspace{-0.7cm}
\end{table}

\subsection{Microbenchmark: availability guarantees}
\label{sec:eval:autozoning-guarantees}

This experiment evaluates to what extent the configuration of localized data is
exposed to remote gray failures.  This time, however, no jurisdictions are
given: we use autozoning and test \name's availability guarantees in comparison
to Physalia.  We ask the question: \say{\textit{If a random site runs a reconfiguration
for a random item, to what extend could remote failures
cause the reconfiguration to fail?}}

\com{
We set up the experiment as follows.  We generate 1000 pairs of writer clients,
each performing a configuration write at a site chosen uniformly at random.
Each pair writes a different key from the other pairs, so as to avoid dependencies across pairs.  As
a result, the reconfiguration RTTs (\ie the RTT between the two sites) are
distributed uniformly at random.
}
%\gf{Alternative:

\methodology{} We generate pairs of writer clients; each pair performs a
configuration write at a site chosen uniformly at random.  The pairs are chosen
as follows. We select 30 random sites, and in a random radius around each site
of up to 150ms chosen uniformly at random, we generate 1000 interacting pairs.
We disconnect the network at a distance $R_12 = x*R_1, x=\{1,2,3,4,5\}$, and
run each experiment separately. We then depict the result of the interaction
(success or fail) relative to the RTT between the interacting nodes and the
distance to the failure.  Each pair writes a different key from the other pairs
to avoid dependencies across pairs.  As a result, the reconfiguration RTTs (\ie
the RTT between the two sites) are distributed uniformly at random.

%}

We are interested in what dependencies the second writer might have on other
sites, and how far in network distance these sites are: Dependencies on other
sites could cause the second writer's configuration write to fail, \eg by
triggering a correlated failure.  For this purpose, we select an area of a
random radius R around the second writer, and we partition the network along
the zone's border.  The two writers are never partitioned from each other,
because by design the second writer has a strong consistency dependency on the
first, and then the second write would fail.  By running this experiment, we
test whether the second writer has dependencies outside the partitioned area.

\Cref{fig:local_limix}, \cref{fig:local_physalia} depict the success or failure
result for each writer-writer pair.  The x axis represents the reconfiguration
RTT, and the y axis represents the network distance to the partitions.  Thanks
to \name's exposure guarantees, reconfigurations in \name succeed more
frequently than in Physalia. Both system register failures below \name's
shield, showing that they do have dependencies nearby. However, Physalia sites
also fail when failures are relatively far: Physalia records failures above
\name's shield because it provides no guarantees for all the pairs interacting
across cells.  Even if two sites are relatively far from failures, if they are
in different cells and their cells are partitioned, their interaction might and
do fail. In contrast, \name autozoning provides a clear availability guarantee,
applicable to all interacting pairs: When failures are farther than the Lamport
exposure bound of the two writers, depicted using the black shield line,
reconfiguration is guaranteed to succeed.

\subsection{Availability under real scenarios}
\label{sec:eval:real}

This experiment, like the previous one, tests to what extent the configuration
of localized data is exposed to remote gray failures, but on a real network and
using realistic trace-based data. Because the AWS testbed has lower RTTs that
are more clustered, Physalia cells are mostly between 10-30ms RTT diameter,
with a single cell up to 50ms.  Our methodology is similar to the experiment
above, with the only exception that the workload of 1000 reconfiguration pairs
of each experiment has a distribution of reconfiguration RTTs matching the one
in trace 1 (\cref{sec:setup}). The workload is global, thus some interacting pairs
cross the partition boundary.

\Cref{fig:aws_limix}, \cref{fig:aws_physalia} plot the results 
only for the non-partitioned interactions.
Our results generally mirror the ones we obtained on the cluster testbed,
with \name outperforming Physalia in almost all tested cases, and for the
same reasons.
There are, however, two notable differences from the previous experiment.
First, \name registers a few failures close to but above the shield. These failures
correspond to a some sites with a more unstable RTT than the others, whereas
we depict the RTT measured at bootstrap time. However, the difference in RTT was minor,
and for all the other sites we observed no violation. Second, we observe a
more pronounced clustering effect of the plotted points, matching roughly our
more clustered topology.

We also summarized the results for each tested workload
as percentage of successes of the maximum possible availability for the non-partitioned
pairs. \cref{fig:best_vs_actual_availability_AWS} shows that \name's
success rate is close to 100\% in most cases, and significantly outperforms Physalia.
We conclude that \name provides strong guarantees on a variety of
testbeds and workloads.

\section{Discussion}
\label{sec:discussion}

\textbf{Limitations.} \name implements the metadata service. While this is an   
important component, it does not by itself limit the Lamport exposure of the    
full service stack.  We leave such dependencies for future work, such as        
upper-layer dependencies (e.g., on Javascript), power grid and network links    
whose failure could violate exposure-limiting policies (e.g., if communication  
between two sites in Germany crosses network links outside of Germany).  This   
limitation could in principle be addresssed via deep structural dependency      
analysis~\cite{zhai14heading}, but it is outside the scope of this paper.

\com{
\textbf{Applying \name to other systems.}
\cb{todo}
\paragraph{\crdb configuration.}
\crdb supports tagging the deployment nodes as belonging to a zone,             
though not to                                                                   
multiple zones, \eg, if a node's tag is ``Germany'', it cannot have any         
other tag, such as ``EU''.                                                      
Even so, we could tag nodes with their smallest region. Then we could           
tag the replicas of the configuration ranges in certain regions.                
However, there fundamental limitation is that the replicas do not migrate with the data,
thus once the data leaves the region, the configuration is no longer collocated.
The leaseholder replica can migrate with the data and speed up configuration    
reads, but configuration writes (\ie reconfigurations) need to reach a majority 
of replicas.                                                                    
A likely \crdb-specific restriction: There seems to be a single configuration range for the entire cluster data.
This means, any region we choose for the configuration cannot be collocated with
its corresponding data, because the data is in different regions. 
}

\section{Related Work}

\cb{TODO: Add the following papers: "Finding Critical Regions and Region-Disjoint Paths in a Network", Stojan Trajanovski et al., https://ieeexplore.ieee.org/abstract/document/6775581}

\cb{TODO: Add
- SkipNet: A Scalable Overlay Network with Practical Locality Properties https://www.microsoft.com/en-us/research/wp-content/uploads/2016/02/tr-2002-92.pdf
- motivated by localized resilience; sort of gets the routing locality bit and even the organizational domain boundaries idea
}

\textbf{CAP tradeoffs.}
Faced with partitions, some systems choose to relax consistency in favor of availability.
Gemini~\cite{li12gemini} distinguishes access types that require a              
strongly- or eventually-consistent reply;
this technique is known as 
segmentation~\cite{gilbertlynch2012cap}. 
Dynamo, a highly-available data store,
takes a similar approach~\cite{dynamo}.
Seredinschi et al.~\cite{seredinschi16icg}                                      
provide the user with several replies, increasing in consistency guarantees,    
enabling the client to perform speculative work.  
In \name, all accesses are    
strongly consistent.

%\cb{add dynamo}
% and the focus is to provide the smallest possible exposure 
%to availability failures and slowdowns.

\textbf{Availability during failures.}
Several strongly                                     consistent systems employ
replication to survive failures and partitions,       however, they assume
uncorrelated failures across sites~\cite{wu13spanstore,lamport01paxos,uluyol20pando,ongaro14raft,vanRenesse04chain}.
Failures across geographical might not be independent, for several reasons:
machines across sites run the same software and are vulnerable to the same
bugs~\cite{haeberlen05glacier,brooker2020millions}; nodes' hard disks fill up
at the same rate~\cite{brooker2020millions}; short-lived and, less frequently,
long-lived (partial) partitions separate sites from each other, causing a
domino of failures and ultimately
unavailability~\cite{alquraan18cloudfailures}.

Unlike the availability metric that Hauer et al.~\cite{hauer20meaningful}       
recently proposed, which \textit{reactively} analyzes failures after they       
occur, \name \textit{proactively} limits exposure in the first place.
\com{In contrast with the \textit{blast   
radius} notion proposed by Brooker et al.~\cite{brooker2020millions}, which     
attempts to reduce damage caused by a partition, \name focuses on users,       
aiming to insulate their accesses from \textit{any} partitions or slowdowns     
outside a relevant local zone.}%
Glacier~\cite{haeberlen05glacier} employs massive replication of data, which
\name also does, to minimize the probability of data loss during large-scale
correlated failures. As opposed to \name, however, Glacier considers only data
stores with immutable objects.

\section{Conclusion}

Can we achieve both the elasticity of globalized computing infrastractures and
the resilience to distant failures of localized infrastructure?  We show that
this is possible by limiting the Lamport exposure and guaranteeing that
any user can access any data at distance $\Delta$ away, when failures occur
beyond a small $O(\log N)$ multiple of $\Delta$.

\com{
\name is an configuration service that provides strong guarantees for  
a user's worst-case availability and performance in the presence of failures    
and partitions. \name satisfies simultaneous bounds for any user accessing any  
item. \name designs a control plane that limits exposure by running a separate  
lookup service per zone so that, if some zones become partitioned and        
unavailable, other ones can respond instead. \name's control plane supports     
administrative zones and existing strongly consistent data planes with item     
migration. But, it also defines efficient and scalable autozoning algorithm     
with tight exposure bounds, and an in-house scalable data plane. These          
techniques together enable \name to achieve up to 100\% better availability 
over \crdb at a logarithmic dynamic overhead.
}

\bibliographystyle{plain}
\bibliography{rnd,new,bib/comp,bib/fault,bib/lang,bib/net,bib/os,bib/priv,bib/sec,bib/soc,bib/theory,hotnets21}

\end{document}